\documentclass[twocolumn,english,aps,prb]{revtex4-1}
\pdfoutput=1
\usepackage[T1]{fontenc}
\usepackage[latin9]{inputenc}
\usepackage[a4paper]{geometry}
\usepackage{amsmath}
\usepackage{amssymb}
\usepackage{graphicx,color}
\usepackage{esint}
\usepackage[titletoc,title]{appendix}
\usepackage{ulem}

\newcommand{\im}{\mathrm{i}}
\newcommand{\e}{\mathrm{e}}

\begin{document}

\title{Channel surface plasmons in a continuous and flat  graphene sheet}
\author{A. J. Chaves}\email{andrej6@gmail.com}
\affiliation{Department of Physics and Center of Physics, and QuantaLab, University of Minho, 4710-057, Braga, Portugal}
\author{D. R. da Costa}\email{diego_rabelo@fisica.ufc.br}
\affiliation{Departamento de F\'isica, Universidade Federal do Cear\'a, Caixa Postal 6030, Campus do Pici, 60455-900 Fortaleza, Cear\'a, Brazil}
\author{G. A. Farias}\email{gil@fisica.ufc.br}
\affiliation{Departamento de F\'isica, Universidade Federal do Cear\'a, Caixa Postal 6030, Campus do Pici, 60455-900 Fortaleza, Cear\'a, Brazil}
\author{N. M. R. Peres}\email{peres@fisica.uminho.pt}
\affiliation{Department of Physics and Center of Physics, and QuantaLab, University of Minho, 4710-057, Braga, Portugal}

\begin{abstract}
We derive an integral equation describing surface-plasmon polaritons in graphene deposited on a substrate with a planar surface and a  dielectric protrusion in the opposite surface of the dielectric slab. We show that the problem is mathematically equivalent to the solution of a Fredholm equation, which we solve exactly. In addition, we show that the dispersion relation of the channel surface plasmons is determined by the geometric parameters of the protrusion alone. We also show that such system supports both even and odd modes. We give the electrostatic potential and the intensity plot of the electrostatic field, which clearly show the transverse localized nature of the surface plasmons in a continuous and flat  graphene sheet.
\end{abstract}
\maketitle

\section{Introduction}

Light-matter interaction at the nanoscale is the realm of plasmonics. In the visible and near-IR spectral range one generally relies on the plasmonic properties of noble metals, such as Gold,  Silver, and Copper. On the other hand, in the mid-infrared (IR) and in the THz the use of noble metals is excluded due to  poor confinement of the
surface plasmons. It is in this context that graphene emerges as a platform for plasmonics in the  mid-IR and in the THz  spectral range, since this material supports strongly confined surface plasmons in this frequency region \cite{ju2011graphene,yan2012tunable}.
 
In the field of plasmonics one can distinguish  between two different types of surface plasmons. For graphene, as for noble metals, the types are: 
(i) surface-plasmon polaritons and (ii) localized surface plasmons. The former
are propagating surface waves on the graphene surface, whereas the latter are localized excitations in graphene nanostructures, such as ribbons \cite{ju2011graphene} and disks\cite{yan2012tunable}.  Thus, in general,
for obtaining localized plasmons in graphene one has to pattern the graphene sheet, which hinders the quality factor of these excitations. It would be, therefore, convenient,
to provide a method to induce localized plasmons in a continuous graphene sheet. To investigate this possibility is the purpose of this paper, where we should the existence
of transversely localized channel plasmons. 

The coupled mode of an electromagnetic field with  charge density oscillations of a
conductor is called surface plasmon-polaritons (SPP). The  synthesis of
graphene and other new two dimensional (2D) materials  opened the door to natural candidates to support this type of surface modes, therefore creating a new
field inside plasmonics \cite{GoncalvesPeres,basov2016polaritons,low2017polaritons}.
The most interesting properties of the SPP are the confinement of light below
the diffraction limit \cite{atwater2007promise}. The  list of applications for
SPP is extensive, including biochemical sensing \cite{brolo2012plasmonics,acimovic2014lspr,rodrigo2015mid}, solar cells \cite{green2012harnessing}, optical tweezers \cite{reece2008plasmonics,juan2011plasmon}, and transformation optics \cite{huidobro2010transformation,pendry2012transformation}.  In planar dielectric-metal interfaces, the SPP is confined along the direction transverse to the surface. However, at dielectric gaps, such as V-shaped groove and nanogaps, such plasmons can be
also be confined along the non translational-invariant direction, being classified as channel plasmon-polaritons (CPP) \cite{SmithRevCPP}. Those systems can be used 
to steer SPP, thus forming   plasmonic waveguides \cite{Nielsen:2008,PileAPL}. 

Inside the sub-field of 2D photonics, graphene attracted much attention due to its unique properties, including the fact that its optical properties can be controlled externaly by  electrostatic gating, originating long-lived plasmons, with large field confinement in the THz and mid-IR spectral range \cite{jablan2009plasmonics,tassin2012comparison,wenger2016optical,GoncalvesPeres,Xiao2016,AbajoACSP}. On the other hand, there are few studies on CPP in graphene. The existing ones discuss
 2D nano-slits\cite{gonccalves2017hybridized} and covered grooves and wedges \cite{Liu:13,Smirnova:2016,gonccalves2017universal}. In all these approaches graphene is deposited in a deformed substrate, thus assuming the same shape of the substrate. This approaches reduce the quality factor of the SPP. Therefore it would be  ideal to 
 find a method where a continuous graphene sheet is deposited on a flat substrate but would still support localized (or channel) plasmons.
 Here we propose a new approach: we take a susbtrate that is flat in one of its surfaces and patterned in the other surface, as can be seen in Fig. \ref{fig:A-one-dimensional-defect}. The graphene sheet is then deposited on the flat region of the substrate, therefore keeping its natural flatness. 
 
Studies of plasmon resonances in a surface with a protuberance or depression were first performed in Ref. \onlinecite{maradudin1985electrostatic}, for the case of noble metals. In Ref. \onlinecite{sturman2014plasmons} localized plasmons in nanoscale pertubations were studied using the  integral equation eigenvalue method, within a quasi-static approximation \cite{mayergoyz2005electrostatic}.  The scattering of SPP by a localized defect in a dielectric with axial symmetry was performed in Ref. \onlinecite{arias2013scattering} using the
reduced Rayleigh equations \cite{zayats2005nano}.
There are numerous works about plasmonic resonances in rough or periodic surfaces, using similar procedures as the one
employed for the  study of a single protuberance \cite{raether1988surface, pereira2003surface,chubchev2017surface}. All these works refer to noble-metal plasmonics and similar geometries have not been yet considered for graphene.

In this paper we calculate, using an electrostatic approximation, the plasmonic transversely-localized modes of a graphene sheet deposited on a flat substrate which has either a protuberance or an
indentation in the opposite face.  The electrostatic approximation is valid when the in-plane momentum $q$ is large compared to the momentum of free radiation inside the media \cite{garcia2013multiple};
 see Ref. \cite{kumar2015tunable} 
 for a comparison between a full electromagnetic calculation with a
electrostatic one for a heterostructure of graphene and hBN.
In Section \ref{sec:2d_protusion} we discuss the procedure to solve a generic 2D dielectric protrusion in the electrostatic approximation and in Section \ref{sec:1d_bump} we derive an
integral equation for the Fourier coefficient of the potential field for a generic 1D deformation. In Section \ref{sec:numerical_solution} we 
discuss the classification of the integral equation, the condition for the existence of transversely-localized plasmons, and the numerical procedure. In Section \ref{sec:results} we discuss our results for a Gaussian profile.

\section{A planar graphene sheet on a dielectric defect: 2D protrusion \label{sec:2d_protusion}}

\begin{figure}
\includegraphics[scale=0.7]{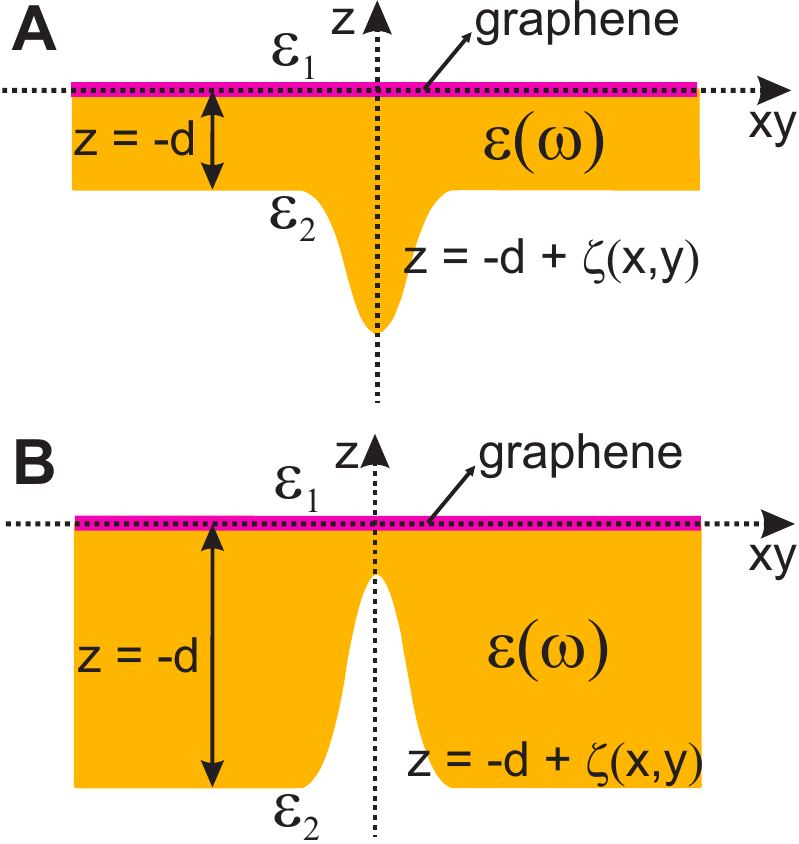}
\caption{\label{fig:A-one-dimensional-defect}\label{fig:A-one-dimensional-defect-groove}A dielectric protrusion 
	(panel A) or indentation (panel B) below a flat
graphene sheet. The defect is assumed to have either even parity symmetry
(1D defect) or cylindrical symmetry (2D defect).}
\end{figure}

Let us consider the geometry of Fig. \ref{fig:A-one-dimensional-defect}.
There are three regions in this system: the region $z>0$, with dielectric
function $\epsilon_{1}$, the region between $-d+\zeta(x,y)<z<0$, with
dielectric function $\epsilon(\omega)$, and the region $-d+z<\zeta(x,y)$,
with dielectric function $\epsilon_{2}$. We have to define the electrostatic
potential in these three regions.

In all regions we write the electrostatic potential as a Fourier integral.
In the first region we write
\begin{equation}
\phi_{1}(\boldsymbol{\rho},z,t)=\int\frac{d\mathbf{k}_{\parallel}}{(2\pi)^{2}}A(\mathbf{k}_{\parallel})\e^{\im\mathbf{k_{\parallel}}\cdot\boldsymbol{\rho}-k_{\parallel}z}\e^{-\im\omega t}\,, \label{eq:region_1}
\end{equation}
where $\mathbf{k}_{\parallel}=(k_x,k_y)$ and $\boldsymbol{\rho}=(x,y)$.
In the central region both real exponentials have to be present, that
is, 
\begin{equation}
\phi_{c}(\boldsymbol{\rho},z,t)=\int\frac{d\mathbf{k}_{\parallel}}{(2\pi)^{2}}[B(\mathbf{k}_{\parallel})\e^{k_{\parallel}z}+C(\mathbf{k}_{\parallel})\e^{-k_{\parallel}z}] \e^{\im\mathbf{k_{\parallel}}\cdot\boldsymbol{\rho}-\im\omega t}. \label{eq:region_c}
\end{equation}
Finally, in the third region we have
\begin{equation}
\phi_{2}(\boldsymbol{\rho},z,t)=\int\frac{d\mathbf{k}_{\parallel}}{(2\pi)^{2}}D(\mathbf{k}_{\parallel})\e^{\im\mathbf{k_{\parallel}}\cdot\boldsymbol{\rho}+k_{\parallel}z}\e^{-\im\omega t}.\label{eq:region_2}
\end{equation}
Next we assume that the above expressions for $\phi_{c}$ and $\phi_{2}$
hold in the  region of bump/protrusion. The boundary conditions are imposed at
$z=0$ and at $z=-d+\zeta(x,y)$, where $\zeta(x,y)$ is some even
function, for example an inverted Gaussian:
\begin{equation}
\zeta(x,y)=-\zeta_{0}\e^{-\rho^{2}/s^{2}},
\end{equation}
where $\rho^{2}=x^{2}+y^{2}$. At $z=0$ the boundary conditions are
the same we have used for solving the flat graphene case discussed in Appendix \ref{app:flat},
whereas for $z=-d+\zeta(x,y)$, the boundary conditions are adapted
from those at $z=0$ considering that $\sigma=0$ (the optical conductivity), and therefore the
normal component of the electric displacement field is continuous
through the interface. Although we have formulated the problem for a defect with cylindrical symmetry, we can also consider one-dimensional profiles,
such as $\zeta(x)=-\zeta_{0}\e^{-4x^{2}/R^{2}}$, which will be the case considered next.

\section{A planar graphene sheet on a dielectric defect: 1D protrusion \label{sec:1d_bump}}


From here on we consider a 1D defect. In this case the Fourier representation
of the field is one-dimensional, reading
\begin{equation}
\phi_{1}(\boldsymbol{\rho},z,t)=\int\frac{dk_{x}}{2\pi}A(k_{x})\e^{\im(k_{x}x+k_{y}y)-k_{\parallel}z}e^{-\im\omega t},
\end{equation}
and similar equations for $\phi_{c}$ and $\phi_{2}$. Next we want
to obtain an eigenvalue equation, thus allowing us to determine the
eigen-frequencies, in terms of a single coefficient. In particular,
we want that coefficient to be $A(k_{x})$. For implementing the boundary
conditions we need an expression for the normal derivative along the
surface of the defect. This is given by
\begin{equation}
\frac{\partial}{\partial n}=\mathbf{\hat{n}}\cdot\nabla=\left[1+\left( \partial_x\zeta(x)\right)^{2}\right]^{-1/2}\left(- \partial_x\zeta(x) \partial_x+ \partial_z\right).
\end{equation}
At the interface $z=0$ we have simply 
\begin{equation}
\frac{\partial}{\partial n}=\frac{\partial}{\partial z}.
\end{equation}
Thus the boundary conditions are
\begin{subequations}
\begin{align}
\phi_{1}(\boldsymbol{\rho},0,t) & =\phi_{c}(\boldsymbol{\rho},0,t),\\
\epsilon_{1}\frac{\partial\phi_{1}(\boldsymbol{\rho},0,t)}{\partial z} & -\epsilon(\omega)\frac{\partial\phi_{c}(\boldsymbol{\rho},0,t)}{\partial z}=-\frac{\im\sigma}{\epsilon_{0}\omega}\nabla_{2D}^{2}\phi(\boldsymbol{\rho},0,t),
\end{align}
\end{subequations}
 which turn into
\begin{subequations}
\begin{align}
A(k_{x}) & =B(k_{x})+C(k_{x}),\\
-\epsilon_{1}A(k_{x}) & -\epsilon(\omega)[B(k_{x})-C(k_{x})]=\kappa A(k_{x}),
\end{align}
\end{subequations}
where we defined:
\begin{equation}
\kappa\equiv\frac{\im\sigma k_{\parallel}}{\epsilon_{0}\omega},
\end{equation}
and the solution for $B(k_{x})$ and $C(k_{x})$ reads
\begin{subequations}
\begin{align}
B(k_{x}) & =A(k_{x})\frac{\epsilon(\omega)-\epsilon_{1}-\kappa}{2\epsilon(\omega)}\equiv A(k_{x})f_{+}(\omega,k_{\parallel}),\label{eq:B_solution}\\
C(k_{x}) & =A(k_{x})\frac{\epsilon(\omega)+\epsilon_{1}+\kappa}{2\epsilon(\omega)}\equiv A(k_{x})f_{-}(\omega,k_{\parallel}). \label{eq:C_solution}
\end{align}
\end{subequations}
This allows us to write the field in the central region as a function
of the $A(k_{x})$ alone. The next step is the implementation of the
boundary conditions at the interface $z_{2c}(x)=-d+\zeta(x)$. These
are
\begin{subequations}
\begin{align}
\phi_{c}(\boldsymbol{\rho},z_{2c}(x),t) & =\phi_{2}(\boldsymbol{\rho},z_{2c}(x),t),\label{eq:Bc2-a}\\
\epsilon(\omega)\frac{\partial\phi_{c}(\boldsymbol{\rho},z_{2c}(x),t)}{\partial n} & =\epsilon_{2}\frac{\partial\phi_{2}(\boldsymbol{\rho},z_{2c}(x),t)}{\partial n}.\label{eq:Bc2-b}
\end{align}
\end{subequations}
The boundary condition (\ref{eq:Bc2-a}) is simply given by 
\begin{align}
\int\frac{dk_{x}}{2\pi}A(k_{x})e^{\im(k_{x}x+k_{y}y)}\left[f_{+}(\omega,k_{\parallel})\e^{k_{\parallel}z_{2c}(x)}+\right.\nonumber\\ \left.+f_{-}(\omega,k_{\parallel})\e^{-k_{\parallel}z_{2c}(x)}\right]
=\int\frac{dk_{x}}{2\pi}D(k_{x})\e^{\im(k_{x}x+k_{y}y)}\e^{k_{\parallel}z_{2c}(x)}\label{eq:Bc2-aa}
\end{align}
The second boundary condition, Eq. (\ref{eq:Bc2-b}), reads
\begin{align}
\epsilon_{2}\int\frac{dk_{x}}{2\pi}D(k_{x})\left[-\frac{\partial\zeta(x)}{\partial x}\im k_{x}+k_{\parallel}\right]\e^{\im(k_{x}x+k_{y}y)}\e^{k_{\parallel}z_{2c}(x)}   =\nonumber \\
\epsilon(\omega)\int\frac{dk_{x}}{2\pi}A(k_{x})\e^{\im (k_{x}x+k_{y}y)}\left[f_{+}(\omega,k_{\parallel})\e^{k_{\parallel}z_{2c}(x)}\times\right.\nonumber\\\times \left.\left(-\frac{\partial\zeta(x)}{\partial x}ik_{x}+k_{\parallel}\right)+f_{-}(\omega,k_{\parallel})\e^{-k_{\parallel}z_{2c}(x)}
\times\right.\nonumber\\ \left. \times\left(-\frac{\partial\zeta(x)}{\partial x}ik_{x}-k_{\parallel}\right)\right]\label{eq:Bc2-bb}
\end{align}
Now we need to combine Eqs. (\ref{eq:Bc2-aa}) and (\ref{eq:Bc2-bb})
for obtaining a single integral equation for the coefficient $A(k_{x})$.
This is a more difficult task since we have the function $z_{2c}(x)$
in the exponent together with derivatives of $\zeta(x)$. For circumventing
this difficulty we introduce the Fourier representation of the exponential
$\e^{\alpha\zeta(x)}$ as
\begin{equation}
\e^{\alpha\zeta(x)}=1+\alpha\int\frac{dQ}{2\pi}J(\alpha;Q)e^{\im Qx},\label{eq:Fourier_of_exponential}
\end{equation}
where
\begin{equation}
J(\alpha;Q)=\int dx\e^{-\im Qx}\frac{\e^{\alpha\zeta(x)}-1}{\alpha}. \label{eq:jota_function}
\end{equation}
Equation (\ref{eq:Fourier_of_exponential}) also implies that 
\begin{equation}
\frac{\partial\zeta(x)}{\partial x}\e^{\alpha\zeta(x)}=\frac{1}{\alpha}\frac{\partial \e^{\alpha\zeta(x)}}{\partial x}=\int\frac{dQ}{2\pi}iQJ(\alpha;Q)\e^{\im Qx}.\label{eq:derivative_of_exponential}
\end{equation}
Eqs. (\ref{eq:Fourier_of_exponential}) and (\ref{eq:derivative_of_exponential})
allow the simplification of Eqs. (\ref{eq:Bc2-aa}) and (\ref{eq:Bc2-bb}).
For eliminating the $D(k_{x})$ coefficient, we multiply Eq.
(\ref{eq:Bc2-aa}) by 
\begin{equation}
\left(\im q_{x}\frac{\partial\zeta(x)}{\partial x}+q_{\parallel}\right)e^{-\im (q_{x}x+q_{y}y)}\e^{q_{\parallel}\zeta(x)},
\end{equation}
where $q_{\parallel}=q=\sqrt{q_{x}^{2}+q_{y}^{2}}$, and multiply
Eq. (\ref{eq:Bc2-bb}) by $\e^{-\im(q_{x}x+q_{Y}y)}\e^{q_{\parallel}\zeta(x)}$. We then  use Eqs. (\ref{eq:Fourier_of_exponential}) and (\ref{eq:derivative_of_exponential}),
and integrate over $\bm{\rho}=(x,y)$. After lengthy calculations
we obtain a single equation involving the coefficient $A(k_{x})$
only:
\begin{equation}
\Gamma(q,\omega)A(q_{x})   = \int_{-\infty}^{\infty} dP \,{\cal K}_{q_y,\omega}(q_x,P) A(P),\label{eq:integral_equation}
\end{equation}
where $\hat{\mathbf{p}}=\mathbf{p}/p$, $\hat{\mathbf{q}}=\mathbf{q}/q$,
$\mathbf{p}=(P,q_{y})$, $\mathbf{q}=(q_{x},q_{y})$, and 
\begin{align}
\Gamma(q,\omega)=[\epsilon_{2}-\epsilon(\omega)]^{-1}\left\{ [\epsilon_{2}-\epsilon(\omega)]\e^{-qd}f_{+}(\omega,q)+\right.\nonumber\\ \left.+[\epsilon_{2}+\epsilon(\omega)]\e^{qd}f_{-}(\omega,q)\right\},
\end{align}
with the Kernel:
\begin{align}
 \,{\cal K}_{q_y,\omega}(q_x,P) =pJ(q-p;q_{x}-P)f_{-}(\omega,p)\e^{pd}(1-\nonumber \\-\hat{\mathbf{p}}\cdot\hat{\mathbf{q}}) 
  - pJ(q+p;q_{x}-P)f_{+}(\omega,p)\e^{-pd}(1+\hat{\mathbf{p}}\cdot\hat{\mathbf{q}}). \label{eq:integral_kernel}
\end{align}

\section{Numerical solution \label{sec:numerical_solution}}

\begin{figure}[h]
\includegraphics[scale=0.19]{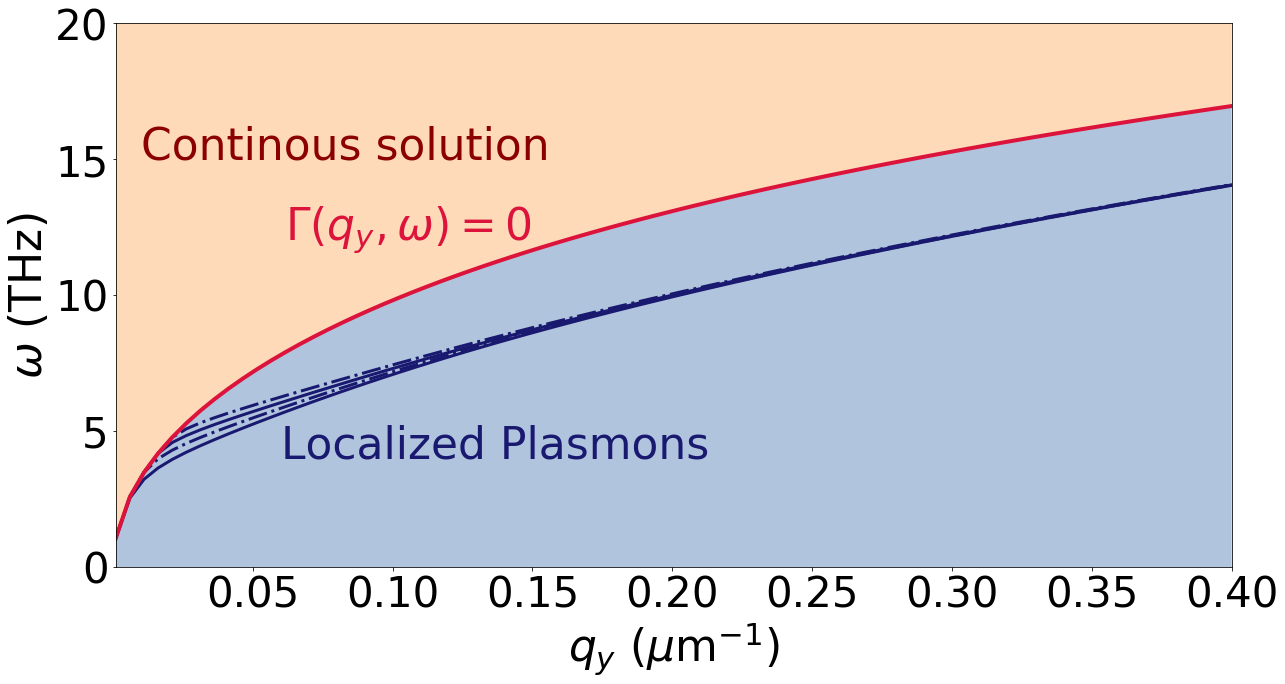}
\caption{Transversely-localized plasmon dispersion. The solid red curve is the solution for $\Gamma(q_y,\omega)=0$. For $\omega>\omega_{spp}$, the system admits continuous solutions, otherwise the solutions are transversely-localized plasmons. The parameters used are: $\varepsilon_1=1.4$, $\varepsilon_2=1$, $\varepsilon=4$, $E_F=0.2$ eV, $d=2\mu$m, $R=250\mu$m, and
$\zeta_0=25\mu$m. The solid(dotted) blue curves are even(odd) solutions. We depict the first four transversely-localized plasmon  modes for the parameters considered.}
\label{fig:dispersion}
\end{figure}

Eq. (\ref{eq:integral_equation}) is a homogeneous integral equation. In the following we discuss 
the classification of the integral equation as a Fredholm equation of second or third kind depending
on the values of $q_y$ and $\omega$. If the equation:
\begin{equation}
\Gamma\left(\sqrt{q_y^2+q_x^2},\omega\right)=0, \label{eq:gamma_zero}
\end{equation}
has real solutions for a real $q_x$, the integral equation (\ref{eq:integral_equation}) is a 
Fredholm equation of the third kind. If not, is a Fredholm equation of the second kind.
The curve $\Gamma\left(q_y,\omega\right)=0$ separates the two regimes (see Fig. \ref{fig:dispersion}): the continuous solution and the localized plasmon states (see next Section for details). 
As we have $\hbar\omega\ll 2E_F$, the interband contributions are negligible. Thus we use the Drude formula for the optical conductivity of graphene without sacrificing accuracy. 

From the curve $\Gamma\left(q_y,\omega\right)=0$ we can test the validity of the electrostatic approximation. Noting that the decaying factor without the electrostatic approximation reads $\kappa_i=\sqrt{q_y^2+q_x^2-\varepsilon_i \omega^2/c^2}$, $i=1,2,3$, we have calculated  for $q_x=0$, $q_y=0.4\,\mu$m$^{-1}$, and $E_F=0.3$ eV that the deviation $1-\kappa_i/q_y$ is no more than $4\%$ in the worst case
	scenario. Therefore, for the parameters used in the figures,
	the electrostatic approximation is justifiable.

\subsection{Continuous solution}

We first assume that Eq. (\ref{eq:gamma_zero}) has solutions for real $q_x=\pm q_x^0$. In this condition
Eq. (\ref{eq:integral_equation})  has not regular solutions, however considering a generalized functional space, the solution has the form  \cite{Bart1973}:
\begin{equation}
A(q_x)= \alpha_1 \delta(q_x-q_x^0)+\alpha_2\delta(q_x+q_x^0)+ A_\text{reg}(q_x), \label{eq:A_continuous}
\end{equation}
where $q_x^0$ is the solution of Eq. (\ref{eq:gamma_zero}) and $A_\text{reg}(q_x)$ is the regular part of the $A(q_x)$. The coefficients $\alpha_1$ and
$\alpha_2$ are determined putting Eq. (\ref{eq:A_continuous}) back to Eq. (\ref{eq:integral_equation}) and making
$q_x=q_x^0$. We can separate the odd and even solutions by making $\alpha_2=\pm\alpha_1$. In the case
that we set the protrusion to zero ($\zeta_0=0$), the regular part $A_\text{reg}(q_x)=0$, and we simply recover the solution for the protrusion-free problem. Eq. (\ref{eq:A_continuous}) can be interpreted as
the sum of the propagating waves from the protrusion-free problem $\alpha_1 \delta(q_x-q_x^0)+\alpha_2\delta(q_x+q_x^0)$ plus a term that comes from the geometric effect of the protrusion.

The integral equation satisfied by the regular part of the solution is obtained substituting  Eq. (\ref{eq:A_continuous}) back to Eq. (\ref{eq:integral_equation}):
\begin{align}
\alpha_1{\cal K}_{q_y,\omega}(q_x,q^0_x)+\alpha_2{\cal K}_{q_y,\omega}(q_x,-q^0_x)+\nonumber\\
+\int_{-\infty}^{\infty} dP \, {\cal K}_{q_y,\omega}(q_x,P)  A_\text{reg}(P)=\Gamma(q,\omega)A_\text{reg}(q_x), \label{eq:integral_regular_aux}
\end{align}
where we used the property obtained by the construction of the function (\ref{eq:A_continuous}):
\begin{equation}
\Gamma\left(\sqrt{q_y^2+q_x^2},\omega\right)A(q_x)=\Gamma\left(\sqrt{q_y^2+q_x^2},\omega\right)A_\text{reg}(q_x).
\end{equation}
Now, making $q_x=\pm q_x^0$, the coefficients $\alpha_i$ satisfy the system of equations:
\begin{subequations}
\begin{align}
\alpha_1{\cal K}_{q_y,\omega}(q^0_x,q^0_x)+\alpha_2{\cal K}_{q_y,\omega}(q^0_x,-q^0_x)+\nonumber\\
+\int_{-\infty}^{\infty} dP \, {\cal K}_{q_y,\omega}(q_x^0,P)  A_\text{reg}(P)=0, \label{eq:integral_regular_aux2}\\
\alpha_1{\cal K}_{q_y,\omega}(-q^0_x,q^0_x)+\alpha_2{\cal K}_{q_y,\omega}(-q^0_x,-q^0_x)+\nonumber\\
+\int_{-\infty}^{\infty} dP \, {\cal K}_{q_y,\omega}(-q_x^0,P)  A_\text{reg}(P)=0, \label{eq:integral_regular_aux3}
\end{align}
\end{subequations}
using the following kernel property ${\cal K}_{q_y,\omega}(q_x,P)={\cal K}_{q_y,\omega}(-q_x,-P)$ 
and that $A_\text{reg}(P)=\pm A_\text{reg}(-P)$ we 
obtain: 
\begin{align}
\int_{-\infty}^{\infty} dP \,\left[{\cal K}_{q_y,\omega}(q_x,P) -{\cal K}_{q_y,\omega}(q^0_x,P)\right] A_\text{reg}(P)=\nonumber\\ =\Gamma(q,\omega)A_\text{reg}(q_x), \label{eq:integral_regular}
\end{align}
so, for a given $q_y$ and $\omega$, one has to solve (\ref{eq:integral_regular})  to obtain
the field in the presence of the protrusion. Note that we have a continuous set of frequencies in this case.

\subsection{Transversely-localized plasmons}

In the case where Eq. (\ref{eq:gamma_zero}) has no real solutions for $q_x$, Eq. (\ref{eq:integral_equation}) is a homogeneous Fredholm equation of the second kind. This equation, for a given $q_y$,
has solutions for some particular values of $\omega$. In the following,
we consider the case where $\varepsilon(\omega)=\varepsilon$, that is, the dielectric function is independent of the frequency. In this case,  Eq. ($\ref{eq:integral_equation}$) 
can be rewritten as:
\begin{equation}
\lambda(\omega){\cal D}_1(q_x,q_y) A={\cal D}_2(q_x,q_y) A, \label{eq:integral_redux}
\end{equation}
where we have the following integral operators:
\begin{subequations}
\begin{align}
{\cal D}_1(q_x,q_y)A=q^2\left[ \frac{\varepsilon_2}{\varepsilon} \sinh(qd)+\cosh(qd)\right]A(q_x)
+\nonumber\\+
\frac{\varepsilon_2-\varepsilon}{2\varepsilon}\int_{-\infty}^{\infty} \frac{dP}{2\pi}p 
\left[J(q+p,q_x-P)\e^{-pd}(q_y^2+pq+\right.\nonumber\\ \left.+Pq_x) +J(q-p,q_x-P)\e^{pd}(q_y^2-pq+Pq_x)\right]A(P), \label{eq:def_D1}
\end{align}
\begin{align}
{\cal D}_2(q_x,q_y)A=\frac{q}{2\varepsilon}\left[ (\varepsilon_2-\varepsilon)(\varepsilon-\varepsilon_1) \e^{-qd}
+ (\varepsilon_2+\varepsilon)(\varepsilon+\right.\nonumber\\ \left.+\varepsilon_1) \e^{+qd}  \right]A(q_x)
 +\frac{\varepsilon_2-\varepsilon}{2\varepsilon}\int_{-\infty}^{\infty} \frac{dP}{2\pi}\left[   (\varepsilon-\varepsilon_1)  \times \right. \nonumber\\ \left. \times J(q+p,q_x-P)\e^{-pd}(q_y^2+pq+Pq_x)
+  \right.\nonumber\\ \left. +(\varepsilon+\varepsilon_1)  J(q-p,q_x-P)\e^{pd}(q_y^2-pq+Pq_x) \right]A(P), \label{eq:def_D2}
\end{align}
\end{subequations}
and we defined:
\begin{equation}
\lambda(\omega)=-\im\frac{\sigma(\omega)}{\varepsilon_0\omega}, \label{eq:def_lambda}
\end{equation}
that has the dimension of length.  To obtain those results we used:
\begin{subequations}
\begin{align}
f_+(\omega,p)=\frac{\varepsilon-\varepsilon_1}{2\varepsilon}+\frac{\lambda(\omega)p}{2\varepsilon},\\
f_-(\omega,p)=\frac{\varepsilon+\varepsilon_1}{2\varepsilon}-\frac{\lambda(\omega)p}{2\varepsilon}.
\end{align}
\end{subequations}

The integral operators ${\cal D}_1$ and ${\cal D}_2 $ do not depend
on the frequency $\omega$. To proceed, we discretize the integrals in Eqs. (\ref{eq:def_D1}) and (\ref{eq:def_D2}). First we apply a cutoff in the momentum $P$: $\int_{-\infty}^\infty\rightarrow
\int_{-\Lambda}^\Lambda $, and $\Lambda$ is chosen to be large enough such that the solution converges (we checked that all boundary conditions are obeyed by the numerical solution). The integral can be discretized by applying Gauss-Legendre quadrature. This will reduce the integral [Eq.
(\ref{eq:integral_redux})] to a generalized eigenvalue problem:
\begin{equation}
\lambda(\omega)\mathbf{D}_1 \mathbf{a}=\mathbf{D}_2 \mathbf{a}, \label{eq:generalized_eig}
\end{equation}
where $\mathbf{D}_{1/2}$ are $N\times N$ matrix, with $N$ the number of Gauss points, and $\mathbf{a}$
is a vector with dimension $N$ that represents the discretized version of the function $A(q_x)$. Solving Eq. (\ref{eq:generalized_eig})  we have the spectrum of eigenvalues $\lambda_n(q_y)$.
 The plasmon frequency then is given by the solution of
\begin{equation}
\lambda(\omega)=\lambda_n(q_y). \label{eq:solution_existence}
\end{equation}
If we assume that the conductivity of graphene is given by the Drude formula:
\begin{equation}
\sigma(\omega)=\sigma_0 \frac{4i}{\pi} \frac{E_F}{\hbar\omega+i\hbar\gamma},
\end{equation}
with $\sigma_0=e^2/(4\hbar)$, $E_F$ the Fermi energy, and $\gamma$ the relaxation rate, the plasmon dispersion is given by:
\begin{equation}
\omega_n(q_y)=\sqrt{4 c \alpha \frac{E_F}{\hbar} \frac{1}{\lambda_n(q_y)}}-i\frac{\gamma}{2}, \label{eq:dispersion_relation}
\end{equation}
with $\alpha=1/137$ the fine structure constant and $c$ the speed of light. We can see from Eq. (\ref{eq:dispersion_relation}) that
the transversely-localized plasmon linewidth is half that of the relaxation rate $\gamma$ in graphene. We have also compared the result obtained from (\ref{eq:dispersion_relation})
with a calculation that also included interband contributions to the optical conductivity and the difference for the frequency
of the surface plasmons  obtained as the solution of Eq. (\ref{eq:solution_existence}) is less than $1\%$.

Recalling the condition for the existence of transversely-localized plasmons, the solution of equation $\Gamma(q_y,\omega)=0$ 
is:
\begin{equation}
\lambda(\omega)=\frac{b(q_y)}{q_y},
\end{equation}
with:
\begin{equation}
b(q_y)= \frac{\varepsilon(\varepsilon_2+\varepsilon_1)+(\varepsilon^2+\varepsilon_1\varepsilon_2)\tanh (q_yd)}{\varepsilon+\varepsilon_2 \tanh (q_yd)},
\end{equation}
and using the definition of $\lambda(\omega)$, Eq. (\ref{eq:def_lambda}):
\begin{equation}
\omega_\text{spp}(q_y)=\sqrt{4 c \alpha \frac{E_F}{\hbar} \frac{q_y}{b(q_y)}}-i\frac{\gamma}{2}.\label{eq:max_freq}
\end{equation}

Using the condition that, for the transversely-localized plasmons, $\lambda_n(q_y)<\lambda(q_y)$, we arrive at
\begin{equation}
\lambda_n(q_y)q_y>b(q_y).
\end{equation}
The latter relation 
defines the region of existence of SPP and 
does not depend on properties of the graphene sheet, that is, it is a purely geometric condition. 
\begin{figure}[h!]
\includegraphics[scale=0.17]{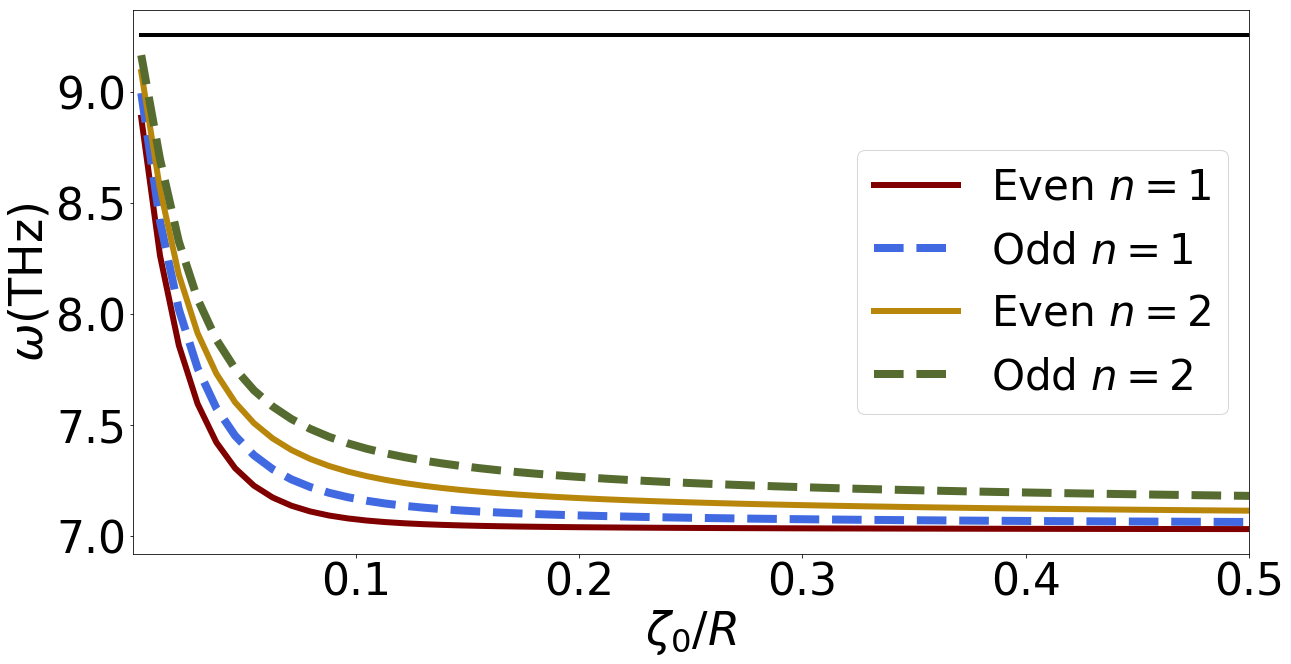}
\caption{Dependence on the ratio $\zeta_0/R$. The black line at $\approx 11$ THz is the 
solution for $\Gamma(q_y,\omega)=0$, i.e., the maximum frequency for transversely-localized plasmons. 
Parameters: $\varepsilon_1=1.4$, $\varepsilon_2=1$, $\varepsilon=4$, $E_F=0.2$ eV, $d=2\mu$m, $R=250\mu$m, $q_y=0.1\mu\mathrm{m}^{-1}$. We clearly see that for $\zeta_0/R\approx 0.2$ the frequency of the transversely-localized plasmon reaches a plateau. } \label{fig:ratio_dependence}
\end{figure}

\begin{figure}[ht]
\includegraphics[scale=0.18]{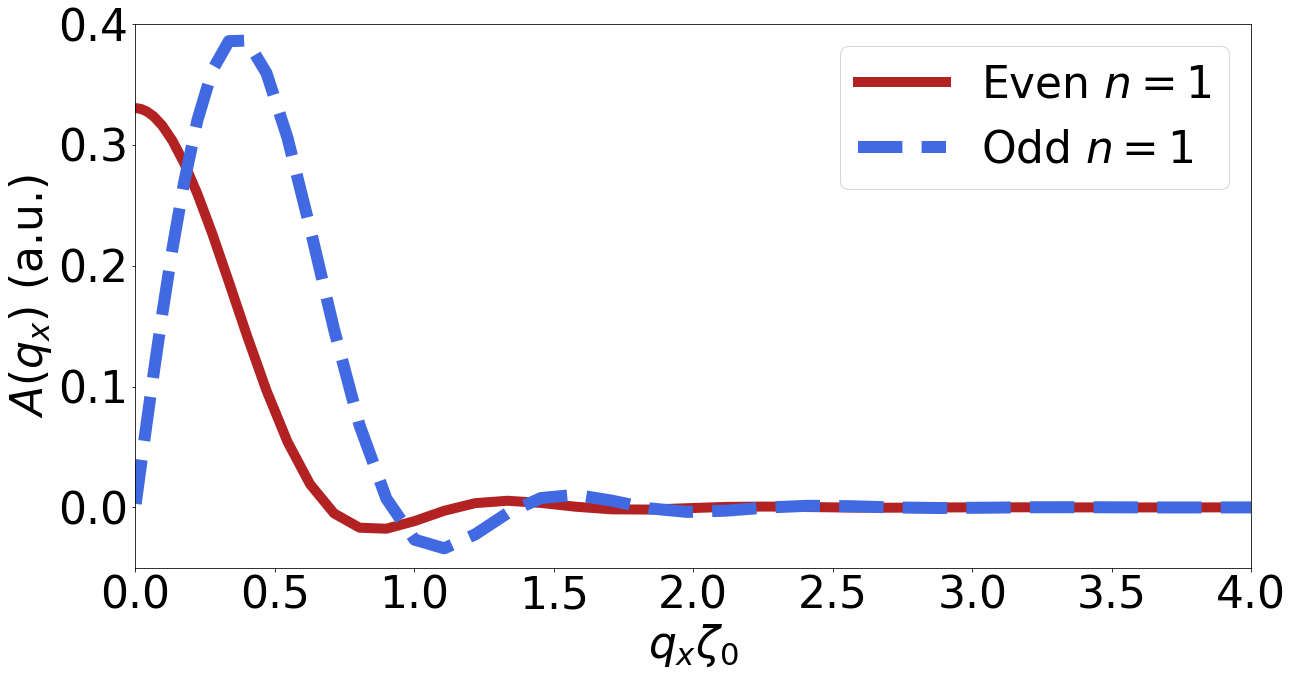}
\caption{Solutions for the generalized eigenvalue problem. The solid red (dashed blue) curve is the first even (odd) solution.   Parameters: $\varepsilon_1=1.4$, $\varepsilon_2=1$, $\varepsilon=4$, $E_F=0.2$ eV, $d=2\mu$m, $R=250\mu$m,
$\zeta_0=25\mu$ m, $q_y=0.4\mu\text{m}^{-1}$. } \label{fig:A_solution}
\end{figure}

Once we compute the coefficient  $A(q_x)$, the coefficients $B(q_x)$ and $C(q_x)$  are calculated using
Eqs. (\ref{eq:B_solution}) and (\ref{eq:C_solution}):
\begin{subequations}
\begin{align}
B(k_{x}) & =A(k_{x})\frac{\epsilon(\omega)-\epsilon_{1}-\kappa}{2\epsilon(\omega)},\label{eq:B_solution2}\\
C(k_{x}) & =A(k_{x})\frac{\epsilon(\omega)+\epsilon_{1}+\kappa}{2\epsilon(\omega)}. \label{eq:C_solution2}
\end{align}
\end{subequations}

The equation for $D(q_x)$ can be obtained from the boundary condition (\ref{eq:Bc2-aa}), using the  same procedure that was used to obtain the Eq. (\ref{eq:integral_equation}): 
\begin{widetext}
\begin{eqnarray}
qe^{-qd}D(q_x)+ \int_{-\infty}^{\infty} \frac{dP}{2\pi} J(q+p,q_x-P)\e^{-pd}(q_y^2+pq+q_xP)D(P)
=q\left(\frac{\varepsilon-\varepsilon_1}{2\varepsilon}\e^{-qd}+\frac{\varepsilon+\varepsilon_1}{2\varepsilon}\e^{qd}-\right. \nonumber \\ \left.-q\lambda(\omega)\frac{\sinh(qd)}{\varepsilon} \right) A(q_x)+\int_{-\infty}^\infty \frac{dP}{2\pi} \left[(q_y^2+pq+Pq_x)f_+(\omega,p)J(q+p,q_x-P)\e^{-pd}+\right.\nonumber\\ \left.
+(q_y^2-pq+Pq_x)f_-(\omega,p)J(q-p,q_x-P)\e^{ pd}\right] A(P),\label{eq:find_d}
\end{eqnarray}
\end{widetext}
which can be written in a matrix from as:
\begin{equation}
\mathbf{G}_1 \mathbf{d}=\mathbf{G}_2 \mathbf{a},
\end{equation}
where $\mathbf{G}_1$ and $\mathbf{G}_2$ are the discrete versions of the operators appearing in
Eq. (\ref{eq:find_d}) and $\mathbf{d}$ is the corresponding
discretized $D(q_x)$ vector. From the previous equation, have the elementary solution:
\begin{equation}
\mathbf{d}=\mathbf{G}_1^{-1}\mathbf{G}_2 \mathbf{a},
\end{equation}
which, since $\mathbf{a}$ has been previously obtained, readily
gives the values for $\mathbf{d}$ by matrix multiplication.
\begin{figure*}[ht]
\includegraphics[scale=0.26]{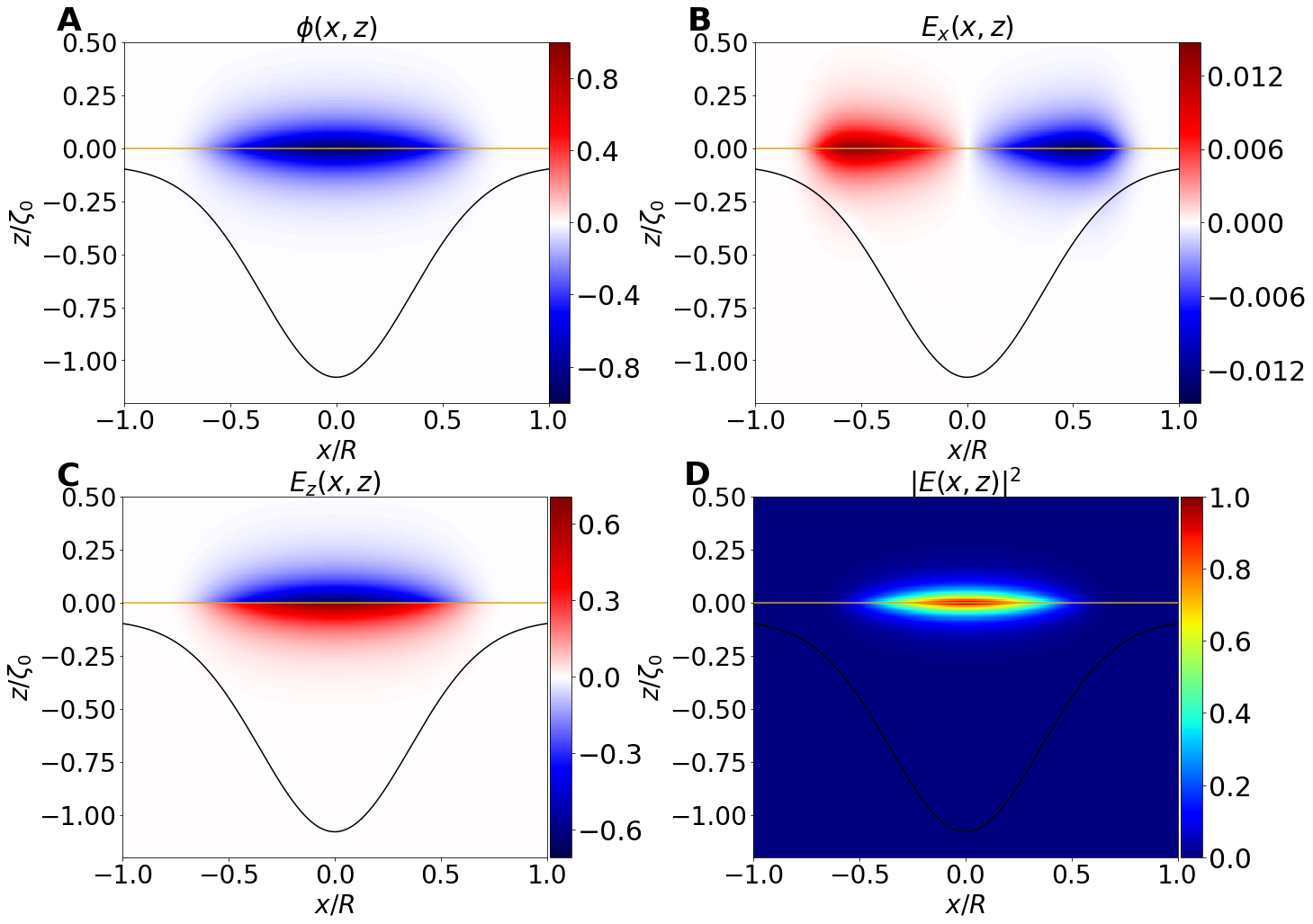}
\caption{Transversely-localized plasmon potential and electric  fields for the system illustrated in Fig \ref{fig:A-one-dimensional-defect}(a). The black line shows the boundary between the regions of dielectric $\varepsilon$ and $\varepsilon_2$ and the gold line represents the graphene sheet. The color represents the intensity of the field in arbitrary units. The electric fields in panels B, C and D are normalized by the maximum value of the total electric field. For the panels B and C the red and blue colors define a change of phase of $\pi$. We show the results for the first even solution. Panels A) Potential Field, B) $x-$ component of the Electric Field, C) $z-$ component of the electric field, D) Square of the absolute value of the electric field. Parameters: $\varepsilon_1=1.4$, $\varepsilon_2=1$, $d=2\mu$m, $R=250\mu$m,
$\zeta_0=25\mu$ m, $q_y=0.4\mu\text{m}^{-1}$. Note that those fields do not depend on the properties of the graphene sheet.}  \label{fig:even_field}
\end{figure*}

\section{Results \label{sec:results}}

From here on, we consider that the protrusion/indentation is described by a Gaussian profile:
\begin{equation}
\zeta(x)=-\zeta_0 e^{-4x^2/R^2},
\end{equation}
with the sign of $\zeta_0$ defining  the two different cases schematically illustrated in Figs. \ref{fig:A-one-dimensional-defect}(A) and \ref{fig:A-one-dimensional-defect}(B). In Appendix \ref{app:gaussian} we calculate the $J(\alpha;Q)$  function [Eq. (\ref{eq:jota_function})] for the Gaussian profile.

From here on, we will consider  the transversely-localized plasmons case only, that is, for a given $q_y$, the maximum 
frequency that we consider is given by Eq. (\ref{eq:max_freq}).  Otherwise specified, we use the following parameters:
$\varepsilon_1=1.4$, $\varepsilon_2=1$, $\varepsilon=4$, $E_F=0.2$ eV, $\gamma=0$, $d=2\,\mu$m, $R=250\,\mu$m, and $\zeta_0=25\,\mu$m. The numerical parameters are $N=100$ Gauss numbers and the cutoff $\Lambda=12/\zeta_0$. These parameters illustrate the implications of the method, but choosing other values amounts to quantitative changes only.

\begin{figure*}[ht]
\includegraphics[scale=0.26]{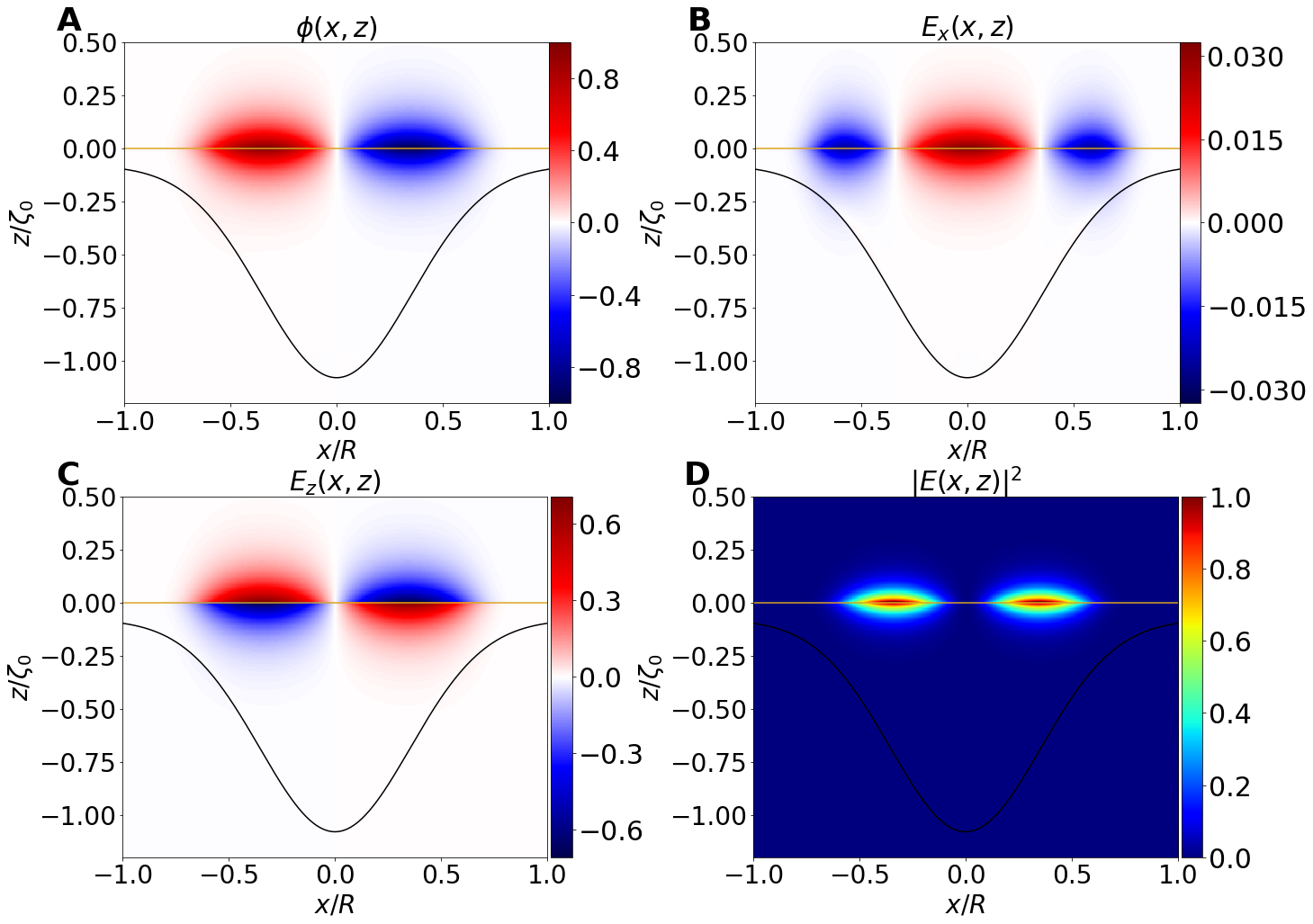}
\caption{The same as Fig. \ref{fig:even_field}, but now for the first odd solution.}
\label{fig:odd_field}
\end{figure*}

\subsection{Parity}

The Kernel of the integral Eq. (\ref{eq:integral_equation}) obeys the identity:
\begin{equation}
{\cal K}(q_x,P)={\cal K}(-q_x,-P),
\end{equation}
and the function $\Gamma(q_x,\omega)$  is  even in the $q_x$ variable. From this condition the solutions
can be classified  in odd and even. Therefore, the limits of integration can be changed as: $\int_{-\Lambda}^\Lambda\rightarrow\int_0^\Lambda$, which simplifies the numerical solution.

\subsection{Scale invariance}

Here we consider how the spectrum changes upon a scale transformation.
Making the scale transformation: $d\rightarrow \xi d$, $R\rightarrow \xi R$, $\zeta_0 \rightarrow \xi\zeta_0$ and $q_y\rightarrow q_y/\xi$ makes the Kernels of Eq. (\ref{eq:integral_redux}) transform Eq. (\ref{eq:def_D1}) to ${\cal D}_1(q_x,q_y) \rightarrow \xi^{-2}{\cal D}_1(q_x/\xi,q_y/\xi) $ and Eq. (\ref{eq:def_D2}) 
${\cal D}_2(q_x,q_y) \rightarrow \xi^{-1}{\cal D}_2(q_x/\xi,q_y/\xi) $ . Therefore the eigenvalue $\lambda_n$ of the matrix equation (\ref{eq:integral_redux}) transform to $\lambda_n\rightarrow \xi \lambda_n $. From Eq. (\ref{eq:dispersion_relation}) the frequency of the transversely-localized plasmon scale as 
$\omega_n(q_y) \rightarrow \xi^{-1}\omega_n(q_y/\xi) $. 
This simple transformation of the eigen-frequencies upon a scale transformation is due to the electrostatic limit we have considered from the outset.

From this discussion, only the ratios $\zeta_0/R$ and $d/R$ matters for the calculation of the dispersion relation. In Fig. \ref{fig:ratio_dependence} we show the dependence on the plasmon frequency for fixed $R$ and $d$ as function of $\zeta_0$, where we can see clearly two regimes:
for  $\zeta_0/R<0.2$, we have a fast change in the localization frequency starting from the continuous solution (maximum localized frequency), and for $\zeta_0/R>0.2$ the system reaches a plateau and the change in the plasmon frequency is negligible.

\subsection{Discussion}

\begin{figure*}
\includegraphics[scale=0.26]{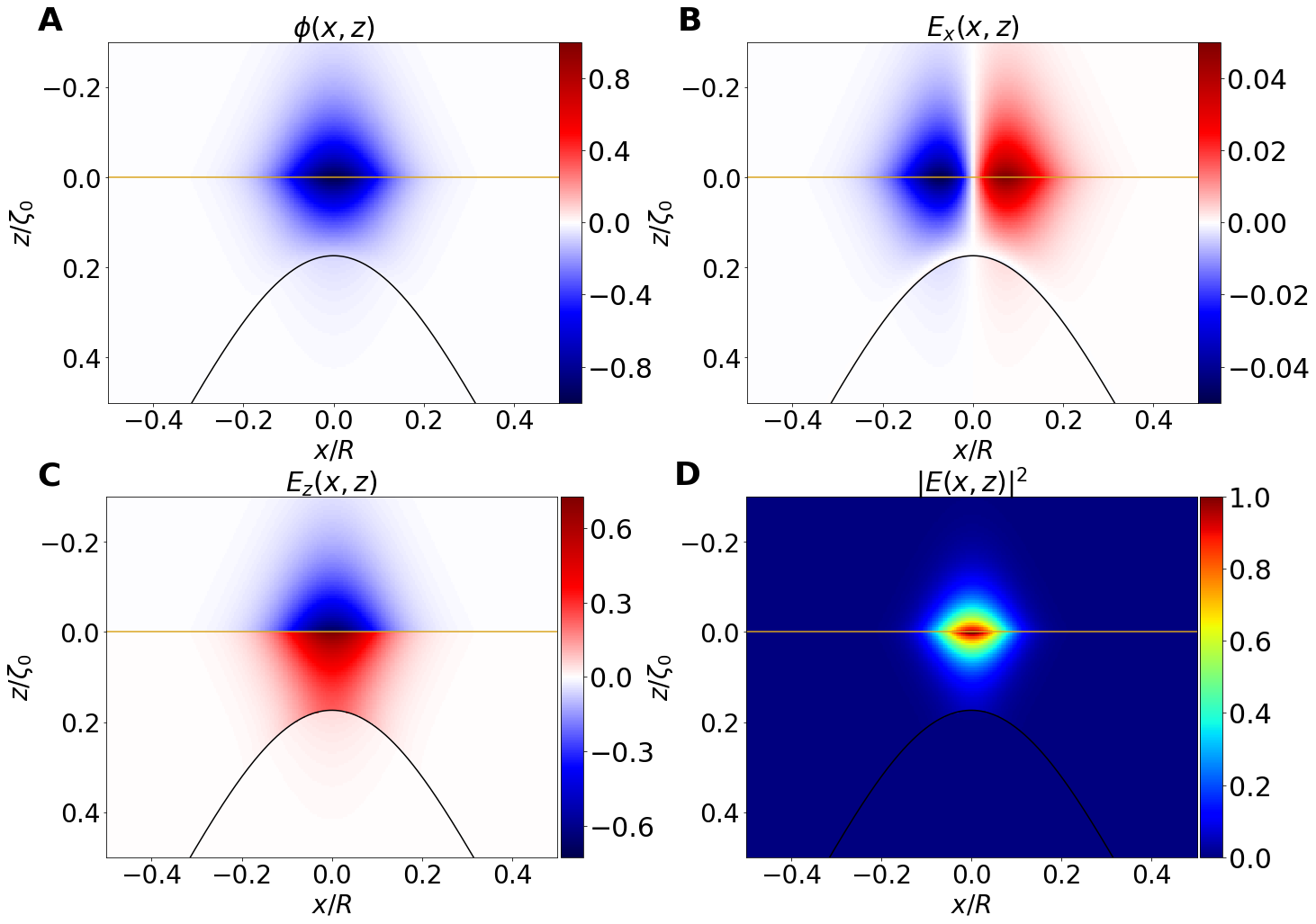}
\caption{Transversely-localized plasmon potential and electric fields for the system illustrated in Fig \ref{fig:A-one-dimensional-defect}(b). The black line shows the boundary between the regions of dielectric $\varepsilon$ and $\varepsilon_2$ and the gold line represents the graphene sheet. The color represents the intensity of the field in arbitrary units. The electric fields in panels B, C and D are normalized by the maximum value of the total electric field. For the panels B and C the red and blue colors define a change of phase of $\pi$. We show the results for the first even solution. Panels A) Potential Field, B) $x-$ component of the Electric Field, C) $z-$ component of the electric field, D) Square of the absolute value of the electric field. Parameters: $\varepsilon_1=1$, $\varepsilon_2=6$, $\varepsilon=1.4$, $d=27\mu$m, $R=250\mu$m,
$\zeta_0=-23\mu$ m, $q_y=0.4\mu\text{m}^{-1}$.}  \label{fig:even_field2}
\end{figure*}

\begin{figure*}
\includegraphics[scale=0.26]{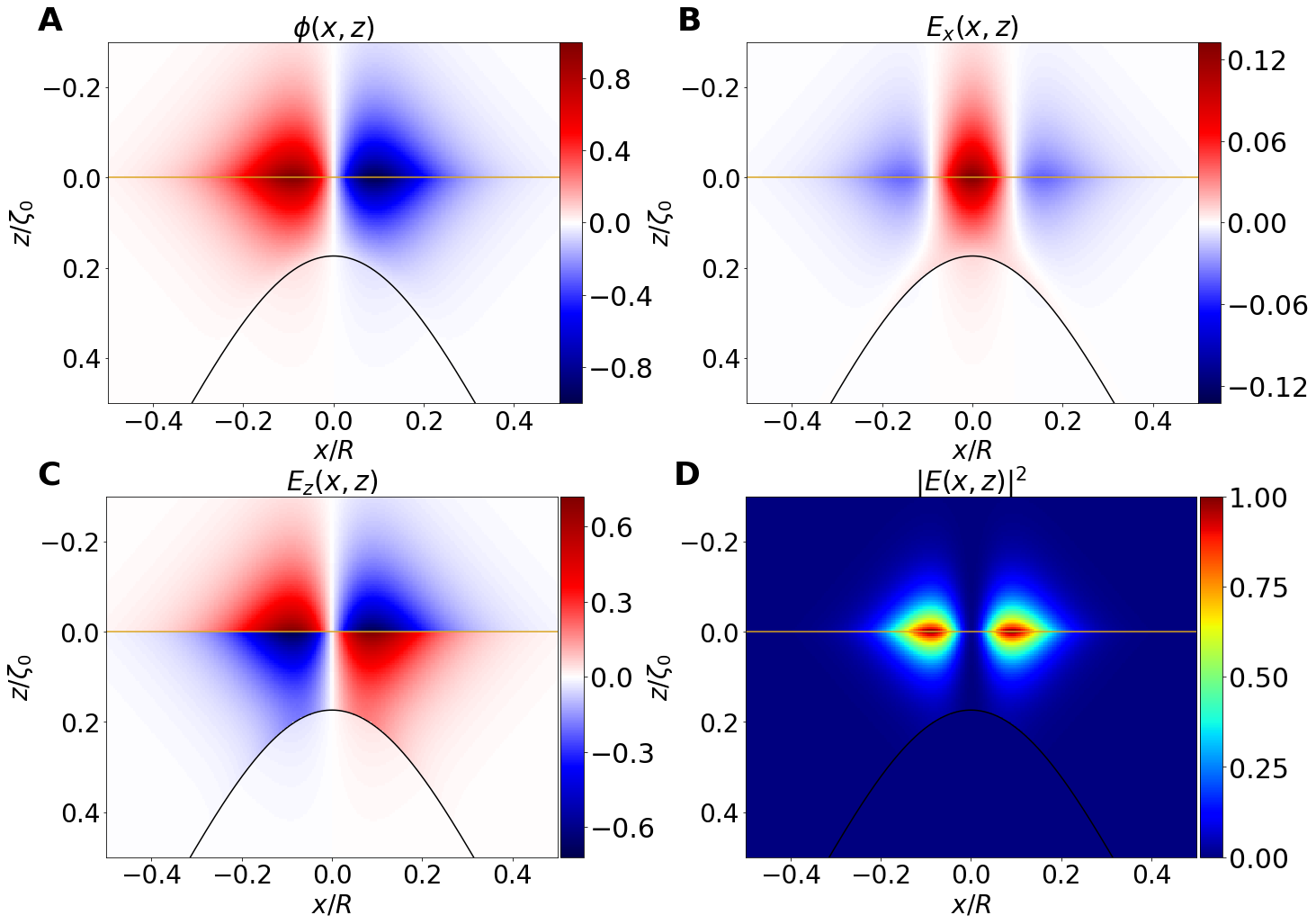}
\caption{The same as Fig. \ref{fig:even_field2}, but now for the first odd solution.}
\label{fig:odd_field2}
\end{figure*}

First we show in Fig. \ref{fig:A_solution} the solution for the generalized eigenvalue problem (\ref{eq:integral_redux}) for the first even and odd solutions and $q_y=0.4\mu\text{m}^{-1}$, where we can see that the functions $A(q_x)$ approach zero for $q_x\zeta_0\approx2$.

Using Eqs. (\ref{eq:B_solution}), (\ref{eq:C_solution}), and (\ref{eq:find_d}) we can compute all the other functions $B(q_x)$, $C(q_x)$,
and $D(q_x)$. The potential field can be calculated now from [see Eqs. (\ref{eq:region_1}),
(\ref{eq:region_c}),and (\ref{eq:region_2})]:
\begin{subequations}
\begin{align}
\phi_{1}(x, 0,z,0) & =\int_{-\infty}^\infty\frac{dP}{2\pi}A(P)\e^{\im Px}\e^{-pz},\label{eq:field_1}\\
\phi_\text{c}(x, 0,z,0) & =\int_{-\infty}^\infty\frac{dP}{2\pi}\e^{\im Px}\left(B(P)\e^{-pz} +C(P)\e^{pz}\right),\label{eq:field_c}\\
\phi_{2}(x, 0,z,0) & =\int_{-\infty}^\infty\frac{dP}{2\pi}D(P)\e^{\im Px}\e^{pz}, \label{eq:field_2}
\end{align}
\end{subequations}
where for simplicity we are only interested for the results to $y=0$ and $t=0$, because of time and $y-$ translation invariance. The electric field can be obtained from $\mathbf{E}=-\boldsymbol{\nabla}\phi$, with $E_{y,i}=-\im k_y \phi_i$, where $i=1,2,\text{c}$ labeling the three regions. The other two components are:
\begin{subequations}
\begin{align}
E_{x,1}(x,y=0,z,t=0) & =-\im \int_{-\infty}^\infty\frac{dP}{2\pi}P A(P)\e^{\im Px}\e^{-pz},\label{eq:field_Ex}\\
E_{z,1}(x,y=0,z,t=0) & =\int_{-\infty}^\infty\frac{dP}{2\pi}p A(P)\e^{\im Px}\e^{-pz},\label{eq:field_Ez}
\end{align}
\end{subequations}
and similar expressions for the regions $2$ and c. 
First we note that from the parity symmetry of the system, the normalization of the field $A(P)$ can be choose such  that $\phi_i$ will be always a real quantity. With a real $\phi_i$, the electric fields $E_{x,i}$ and $E_{z,i}$ will also be real and $E_{y,i}$ will be a pure imaginary quantity, i.e., it will always be out-of-phase by $\pi/2$ with the other electric field components.

From Eqs.  (\ref{eq:field_1})--(\ref{eq:field_Ez}) we calculate the potential and electrical field in Fig. \ref{fig:even_field},  for the first even solution and in Fig. \ref{fig:odd_field} 
for the first odd solution in a Gaussian 1D protrusion. For those solutions the  plasmon frequencies are $\omega_\text{even}=14.04$ THz and $\omega_\text{odd}=14.05$ THz, respectively. The even solution has a node at $x=0$, as it should be, and the field strength, as can be seen in the panel D, is concentrated in the $x$ axis for $x\approx 0.25R=60\mu m$ and in
the $y$ axis around $y\approx0.1\zeta_0\approx2.5\mu m$, far below the  wavenumber $\lambda=134 \mu$m for the light in air. 
 We have verified that the fields obtained by Eqs. (\ref{eq:field_1})--(\ref{eq:field_Ez}) satisfy all the boundary conditions. In Figs. \ref{fig:even_field2} and \ref{fig:odd_field2} we show the transversely-localized plasmons in a groove (Gaussian indentation), where we can see that the field is less localized in the $y$ axis
in comparison with  the protrusion case. However, the indentation ``squeezes" the plasmon in the central region.
 A remarkable characteristic of the electrostatic approximation is that all the fields profile are only a geometric solution of the integral equation, that is, they do not depend on the properties of the graphene sheet. However, they can only exist if Eq. (\ref{eq:solution_existence}) has solution. Therefore, without the graphene sheet there are no transversely-localized plasmons.

We also note that the Fermi-energy $E_F$ can be used to tune the frequency of the transversely-localized plasmons as per 
Eq. (\ref{eq:dispersion_relation}). Another important result of
our study is that the transversely-localized plasmon dispersion is always below the usual SPP dispersion (see Fig. \ref{fig:dispersion}).
Therefore,  for a given frequency, the wavelength of the transversely-localized plasmons is always smaller than its value for a continuous graphene sheet on a homogeneous dielectric, implying a higher degree of confinement of the 
plasmons in dielectric with a protrusion/indentation.

\begin{figure}
\includegraphics[scale=0.17]{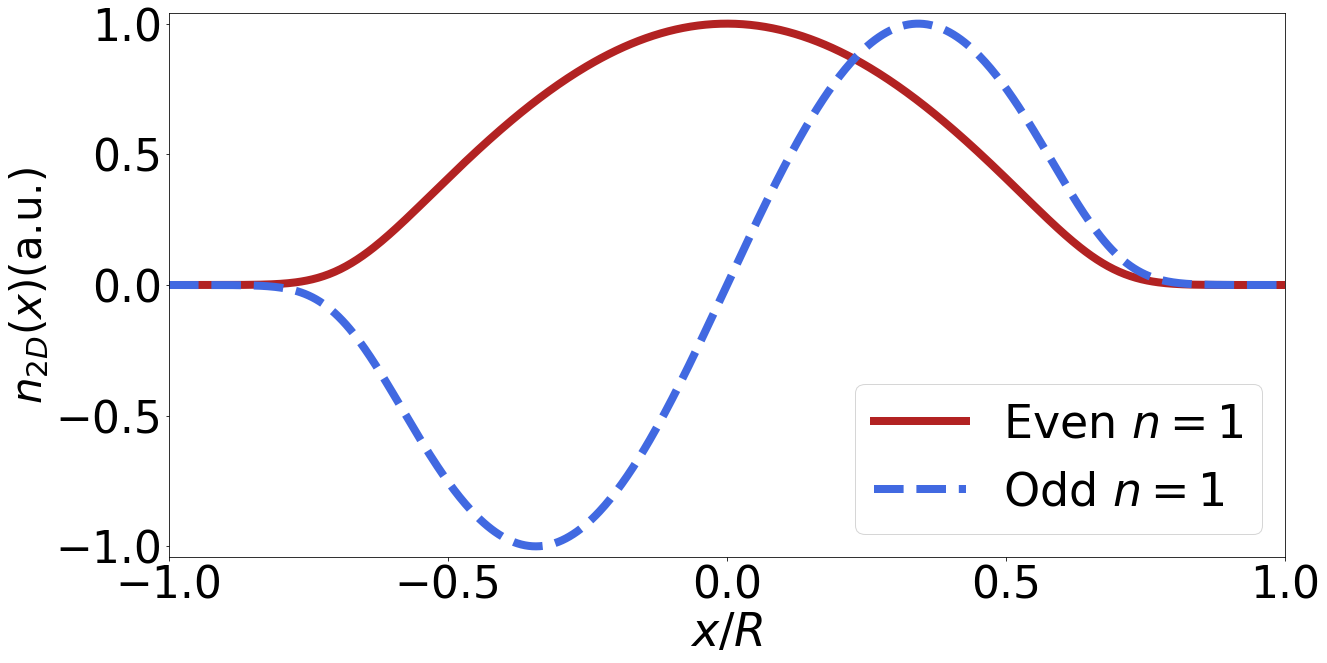}
\caption{Charge density at the graphene sheet.  Solid red (dashed blue) curve is the first even (odd) solution. Parameters: $\varepsilon_1=1.4$, $\varepsilon_2=1$, $\varepsilon=4$, $E_F=0.2$ eV, $d=2\mu$m, $R=250\mu$m,
$\zeta_0=25\mu$ m, $q_y=0.4\mu\text{m}^{-1}$.} \label{fig:charge_density}
\end{figure}

The charge density can be calculated using the equivalent of Eq. (\ref{eq:charge_density}):
\begin{equation}
n_\text{2D}(x)=-\lambda(\omega)\int_{-\infty}^\infty\frac{dP}{2\pi}(P^2+k_y^2) A(P)\e^{\im Px},
\end{equation}
and we show the charge density for the first two modes in the groove in Fig. \ref{fig:charge_density},
where we can see again that the charge is localized around $x\approx0.5R$ the center of the wedge.
Finally, we note that we have used a Gaussian profile but our approach can be used to any differentiable profile.

\section{Conclusions}\label{sec:conclusions}

In this paper we have developed an approach of creating transversely localized plasmons
in a flat graphene sheet. This is possible in a configuration where 
graphene rests  on a flat substrate
with the opposite surface  of the latter showing a protrusion or and indentation (a defect). 
The transversely-localized plasmons dispersion relation appears below the dispersion relation of the 
propagating plasmons when graphene rests on
a flat dielectric of thickness $d$. Above this latter dispersion relation, we have found a
continuum of states, which would be needed for describing scattering by the defect.
Therefore, we have shown that a defect (in this case with even symmetry) can trap localized 
surface plasmons. Since the defect is 1D, the wave number along the axis of symmetry of the 
defect is well defined and, therefore, this defect can also act as channel for propagation of the 
transversely-localized surface plasmons. This geometry has the advantage  of being 
 unnecessary to pattern the graphene sheet, therefore it works without deteriorating the electronic mobility of  graphene. The generalization of the problem dealt in this paper to a 2D defect 
 is straightforward, involving only extra computer power. This is no impediment as our codes 
 are fast enough and run in a laptop in only few minutes. 

\begin{acknowledgments}
	A.J.C. acknowledges for a scholarship from the Brazilian agency CNPq
	(Conselho Nacional de Desenvolvimento Cient\'ifico e Tecnol{\'o}gico).
N.M.R.P. acknowledges  the European Commission through
the project ``Graphene-Driven Revolutions in ICT and Beyond\textquotedbl{}
(Ref. No. 696656) and the Portuguese Foundation for Science and Technology
(FCT) in the framework of the Strategic Financing UID/FIS/04650/2013. 
D. R. C. and G. A. F. acknowledges CNPq under the PRONEX/FUNCAP grants and the CAPES
Foundation.
\end{acknowledgments}

\appendix

\section{A planar graphene sheet}\label{app:flat}

Let us assume a graphene sheet located at $z=0$ in the $xy-$plane.
The graphene is capped by two dielectrics with dielectric functions
$\epsilon_{1}$, for $z>0$, and $\epsilon_{2}$, for $z<0$. We want
to find the spectrum of graphene SPP. The solution of Laplace's equation
for $z>0$ reads
\begin{equation}
\phi_{1}(\mathbf{\bm{\rho}},z,t)=A_{1}e^{i\mathbf{k}_{\parallel}\cdot\mathbf{\bm{\rho}}}e^{-k_{\parallel}z}e^{-i\omega t}\equiv\phi_{1}(\bm{\rho},z,\omega)e^{-i\omega t},
\end{equation}
where $\mathbf{k}_{\parallel}=(k_{x},k_{y})$ and $\bm{\rho}=(x,y)$,
and for $z<0$ it is given by 
\begin{equation}
\phi_{2}(\mathbf{\bm{\rho}},z,t)=A_{2}e^{i\mathbf{k}_{\parallel}\cdot\mathbf{\bm{\rho}}}e^{k_{\parallel}z}e^{-i\omega t}\equiv\phi_{2}(\bm{\rho},z,\omega)e^{-i\omega t}.
\end{equation}

The boundary conditions are
\begin{subequations}
\begin{align}
\phi_{1}(\bm{\rho},0,t) & =\phi_{2}(\bm{\rho},0,t)\\
\epsilon_{1}\frac{\partial\phi_{1}(\bm{\rho},0,t)}{\partial n} & -\epsilon_{2}\frac{\partial\phi_{2}(\bm{\rho},0,t)}{\partial n}=-\frac{n_{2D}(\bm{\rho},0,t)}{\epsilon_{0}},
\end{align}
\end{subequations}
where $n_{2D}(\bm{\rho},0,t)$ is the charge density in graphene,
whose time dependence can be explicitly made as $n_{2D}(\bm{\rho},0,t)=n_{2D}(\bm{\rho},0,\omega)e^{-i\omega t}$.
The first boundary condition expresses the continuity of the electrostatic
potential and the second one the discontinuity of the normal component
of the displacement vector. In addition, the electronic density obeys
the continuity equation in frequency space: $i\omega n_{2D}(\bm{\rho},0,\omega)=\nabla_{2D}\cdot\mathbf{J}_{2D}(\bm{\rho},0,\omega)$,
where $\nabla_{2D}=(\partial/\partial_{x},\partial/\partial_{y})$.
Since the electric current density obeys Ohm's law, $\mathbf{J}_{2D}(\bm{\rho},0,\omega)=-\sigma\nabla_{2D}\phi(\bm{\rho},0,\omega)$,
it follows that 
\begin{align}
i\omega n_{2D}(\bm{\rho},0,\omega) & =-\sigma\nabla_{2D}^{2}\phi(\bm{\rho},0,\omega)=\sigma k_{\parallel}^{2}\phi(\bm{\rho},0,\omega). \label{eq:charge_density}
\end{align}
Finally, we have for the 2D electronic density the result
\begin{equation}
n_{2D}(\bm{\rho},0,\omega)=-\frac{i\sigma}{\omega}k_{\parallel}^{2}\phi(\bm{\rho},0,\omega)
\end{equation}
The first boundary condition implies $A_{1}=A_{2}$ and the second
boundary condition gives
\begin{equation}
-\epsilon_{1}k_{\parallel}-\epsilon_{2}k_{\parallel}=\frac{i\sigma}{\omega\epsilon_{0}}k_{\parallel}^{2}
\end{equation}
or
\begin{equation}
\frac{\epsilon_{1}}{k_{\parallel}}+\frac{\epsilon_{2}}{k_{\parallel}}+\frac{i\sigma}{\omega\epsilon_{0}}=0,
\end{equation}
which is the condition giving the dispersion relation of the SPP in graphene, Note that for $\sigma=0$ we recover
the condition giving the dispersion of SPP
at the interface between two dielectrics. In general, we should have
written the electrostatic potential as 
\begin{align}
\phi_{1}(\boldsymbol{\rho},z,t) & =\int\frac{d\mathbf{k}_{\parallel}}{(2\pi)^{2}}A_{1}(\mathbf{k}_{\parallel})e^{i\mathbf{k}_{\parallel}\cdot\mathbf{\boldsymbol{\rho}}}e^{-k_{\parallel}z}e^{-i\omega t},
\end{align}
 and an identical expression for $\phi_{2}(\bm{\rho},t)$, except for
the dependence $e^{-k_{\parallel}z}$ which should by replaced by
$e^{k_{\parallel}z}$. This way of writing the electrostatic potential
is appropriate for discussing rough surface and defects.

\section{The case of a Gaussian profile} \label{app:gaussian}

In this appendix we give the evaluation of the function $J(\alpha;Q)$ for
the Gaussian profile. This is accomplished by expanding the exponential
of $\zeta(x)$ in the integrand, that is, 
\begin{align}
J(\alpha;Q)&=\int_{-\infty}^{\infty}dx\e^{-\im Qx}\frac{\e^{\alpha\zeta(x)}-1}{\alpha}=\nonumber\\
&=\sum_{n=1}^{\infty}\int_{-\infty}^{\infty}dx\e^{-\im Qx}\frac{\alpha^{n-1}}{n!}\zeta^{n}(x).
\end{align}
Therefore we need to compute the integral (the Fourier transform of
a Gaussian)
\begin{align}
I(n;Q)&=(-\zeta_{0})^{n}\int_{-\infty}^{\infty}dx\e^{-\im Qx}\e^{-4nx^{2}/R^{2}}\nonumber\\
&=(-\zeta_{0})^{n}\frac{R\sqrt{\pi}}{2\sqrt{n}}\e^{-Q^{2}R^{2}/(16n)}.
\end{align}
We have then
\begin{equation}
J(\alpha;Q)=\frac{R^{2}\sqrt{\pi}}{2}\sum_{n=1}^{\infty}\frac{(\alpha R)^{n-1}}{\sqrt{n}n!}(-\zeta_{0}/R)^{n}\e^{-Q^{2}R^{2}/(16n)},
\end{equation}
which is a purely geometric quantity.

\bibliographystyle{apsrev4-1}

\begin{thebibliography}{37}%
	\makeatletter
	\providecommand \@ifxundefined [1]{%
		\@ifx{#1\undefined}
	}%
	\providecommand \@ifnum [1]{%
		\ifnum #1\expandafter \@firstoftwo
		\else \expandafter \@secondoftwo
		\fi
	}%
	\providecommand \@ifx [1]{%
		\ifx #1\expandafter \@firstoftwo
		\else \expandafter \@secondoftwo
		\fi
	}%
	\providecommand \natexlab [1]{#1}%
	\providecommand \enquote  [1]{``#1''}%
	\providecommand \bibnamefont  [1]{#1}%
	\providecommand \bibfnamefont [1]{#1}%
	\providecommand \citenamefont [1]{#1}%
	\providecommand \href@noop [0]{\@secondoftwo}%
	\providecommand \href [0]{\begingroup \@sanitize@url \@href}%
	\providecommand \@href[1]{\@@startlink{#1}\@@href}%
	\providecommand \@@href[1]{\endgroup#1\@@endlink}%
	\providecommand \@sanitize@url [0]{\catcode `\\12\catcode `\$12\catcode
		`\&12\catcode `\#12\catcode `\^12\catcode `\_12\catcode `\%12\relax}%
	\providecommand \@@startlink[1]{}%
	\providecommand \@@endlink[0]{}%
	\providecommand \url  [0]{\begingroup\@sanitize@url \@url }%
	\providecommand \@url [1]{\endgroup\@href {#1}{\urlprefix }}%
	\providecommand \urlprefix  [0]{URL }%
	\providecommand \Eprint [0]{\href }%
	\providecommand \doibase [0]{http://dx.doi.org/}%
	\providecommand \selectlanguage [0]{\@gobble}%
	\providecommand \bibinfo  [0]{\@secondoftwo}%
	\providecommand \bibfield  [0]{\@secondoftwo}%
	\providecommand \translation [1]{[#1]}%
	\providecommand \BibitemOpen [0]{}%
	\providecommand \bibitemStop [0]{}%
	\providecommand \bibitemNoStop [0]{.\EOS\space}%
	\providecommand \EOS [0]{\spacefactor3000\relax}%
	\providecommand \BibitemShut  [1]{\csname bibitem#1\endcsname}%
	\let\auto@bib@innerbib\@empty
	\bibitem [{\citenamefont {Ju}\ \emph {et~al.}(2011)\citenamefont {Ju},
		\citenamefont {Geng}, \citenamefont {Horng}, \citenamefont {Girit},
		\citenamefont {Martin}, \citenamefont {Hao}, \citenamefont {Bechtel},
		\citenamefont {Liang}, \citenamefont {Zettl}, \citenamefont {Shen} \emph
		{et~al.}}]{ju2011graphene}%
	\BibitemOpen
	\bibfield  {author} {\bibinfo {author} {\bibfnamefont {L.}~\bibnamefont
			{Ju}}, \bibinfo {author} {\bibfnamefont {B.}~\bibnamefont {Geng}}, \bibinfo
		{author} {\bibfnamefont {J.}~\bibnamefont {Horng}}, \bibinfo {author}
		{\bibfnamefont {C.}~\bibnamefont {Girit}}, \bibinfo {author} {\bibfnamefont
			{M.}~\bibnamefont {Martin}}, \bibinfo {author} {\bibfnamefont
			{Z.}~\bibnamefont {Hao}}, \bibinfo {author} {\bibfnamefont {H.~A.}\
			\bibnamefont {Bechtel}}, \bibinfo {author} {\bibfnamefont {X.}~\bibnamefont
			{Liang}}, \bibinfo {author} {\bibfnamefont {A.}~\bibnamefont {Zettl}},
		\bibinfo {author} {\bibfnamefont {Y.~R.}\ \bibnamefont {Shen}},  \emph
		{et~al.},\ }\href@noop {} {\bibfield  {journal} {\bibinfo  {journal} {Nat.
				Nanotechnol.}\ }\textbf {\bibinfo {volume} {6}},\ \bibinfo {pages} {630}
		(\bibinfo {year} {2011})}\BibitemShut {NoStop}%
	\bibitem [{\citenamefont {Yan}\ \emph {et~al.}(2012)\citenamefont {Yan},
		\citenamefont {Li}, \citenamefont {Chandra}, \citenamefont {Tulevski},
		\citenamefont {Wu}, \citenamefont {Freitag}, \citenamefont {Zhu},
		\citenamefont {Avouris},\ and\ \citenamefont {Xia}}]{yan2012tunable}%
	\BibitemOpen
	\bibfield  {author} {\bibinfo {author} {\bibfnamefont {H.}~\bibnamefont
			{Yan}}, \bibinfo {author} {\bibfnamefont {X.}~\bibnamefont {Li}}, \bibinfo
		{author} {\bibfnamefont {B.}~\bibnamefont {Chandra}}, \bibinfo {author}
		{\bibfnamefont {G.}~\bibnamefont {Tulevski}}, \bibinfo {author}
		{\bibfnamefont {Y.}~\bibnamefont {Wu}}, \bibinfo {author} {\bibfnamefont
			{M.}~\bibnamefont {Freitag}}, \bibinfo {author} {\bibfnamefont
			{W.}~\bibnamefont {Zhu}}, \bibinfo {author} {\bibfnamefont {P.}~\bibnamefont
			{Avouris}}, \ and\ \bibinfo {author} {\bibfnamefont {F.}~\bibnamefont
			{Xia}},\ }\href@noop {} {\bibfield  {journal} {\bibinfo  {journal} {Nat.
				Nanotechnol.}\ }\textbf {\bibinfo {volume} {7}},\ \bibinfo {pages} {330}
		(\bibinfo {year} {2012})}\BibitemShut {NoStop}%
	\bibitem [{\citenamefont {Gon\c{c}alves}\ and\ \citenamefont
		{Peres}(2016)}]{GoncalvesPeres}%
	\BibitemOpen
	\bibfield  {author} {\bibinfo {author} {\bibfnamefont {P.~A.~D.}\
			\bibnamefont {Gon\c{c}alves}}\ and\ \bibinfo {author} {\bibfnamefont
			{N.~M.~R.}\ \bibnamefont {Peres}},\ }\href {\doibase 10.1142/9948} {\emph
		{\bibinfo {title} {An Introduction to Graphene Plasmonics}}}\ (\bibinfo
	{publisher} {World Scientific},\ \bibinfo {address} {Singapore},\ \bibinfo
	{year} {2016})\BibitemShut {NoStop}%
	\bibitem [{\citenamefont {Basov}\ \emph {et~al.}(2016)\citenamefont {Basov},
		\citenamefont {Fogler},\ and\ \citenamefont
		{de~Abajo}}]{basov2016polaritons}%
	\BibitemOpen
	\bibfield  {author} {\bibinfo {author} {\bibfnamefont {D.}~\bibnamefont
			{Basov}}, \bibinfo {author} {\bibfnamefont {M.}~\bibnamefont {Fogler}}, \
		and\ \bibinfo {author} {\bibfnamefont {F.~G.}\ \bibnamefont {de~Abajo}},\
	}\href@noop {} {\bibfield  {journal} {\bibinfo  {journal} {Science}\ }\textbf
		{\bibinfo {volume} {354}},\ \bibinfo {pages} {aag1992} (\bibinfo {year}
		{2016})}\BibitemShut {NoStop}%
	\bibitem [{\citenamefont {Low}\ \emph {et~al.}(2017)\citenamefont {Low},
		\citenamefont {Chaves}, \citenamefont {Caldwell}, \citenamefont {Kumar},
		\citenamefont {Fang}, \citenamefont {Avouris}, \citenamefont {Heinz},
		\citenamefont {Guinea}, \citenamefont {Martin-Moreno},\ and\ \citenamefont
		{Koppens}}]{low2017polaritons}%
	\BibitemOpen
	\bibfield  {author} {\bibinfo {author} {\bibfnamefont {T.}~\bibnamefont
			{Low}}, \bibinfo {author} {\bibfnamefont {A.}~\bibnamefont {Chaves}},
		\bibinfo {author} {\bibfnamefont {J.~D.}\ \bibnamefont {Caldwell}}, \bibinfo
		{author} {\bibfnamefont {A.}~\bibnamefont {Kumar}}, \bibinfo {author}
		{\bibfnamefont {N.~X.}\ \bibnamefont {Fang}}, \bibinfo {author}
		{\bibfnamefont {P.}~\bibnamefont {Avouris}}, \bibinfo {author} {\bibfnamefont
			{T.~F.}\ \bibnamefont {Heinz}}, \bibinfo {author} {\bibfnamefont
			{F.}~\bibnamefont {Guinea}}, \bibinfo {author} {\bibfnamefont
			{L.}~\bibnamefont {Martin-Moreno}}, \ and\ \bibinfo {author} {\bibfnamefont
			{F.}~\bibnamefont {Koppens}},\ }\href@noop {} {\bibfield  {journal} {\bibinfo
			{journal} {Nature materials}\ }\textbf {\bibinfo {volume} {16}},\ \bibinfo
		{pages} {182} (\bibinfo {year} {2017})}\BibitemShut {NoStop}%
	\bibitem [{\citenamefont {Atwater}(2007)}]{atwater2007promise}%
	\BibitemOpen
	\bibfield  {author} {\bibinfo {author} {\bibfnamefont {H.~A.}\ \bibnamefont
			{Atwater}},\ }\href@noop {} {\bibfield  {journal} {\bibinfo  {journal}
			{Scientific American}\ }\textbf {\bibinfo {volume} {296}},\ \bibinfo {pages}
		{56} (\bibinfo {year} {2007})}\BibitemShut {NoStop}%
	\bibitem [{\citenamefont {Brolo}(2012)}]{brolo2012plasmonics}%
	\BibitemOpen
	\bibfield  {author} {\bibinfo {author} {\bibfnamefont {A.~G.}\ \bibnamefont
			{Brolo}},\ }\href@noop {} {\bibfield  {journal} {\bibinfo  {journal} {Nature
				Photonics}\ }\textbf {\bibinfo {volume} {6}},\ \bibinfo {pages} {709}
		(\bibinfo {year} {2012})}\BibitemShut {NoStop}%
	\bibitem [{\citenamefont {Acimovic}\ \emph {et~al.}(2014)\citenamefont
		{Acimovic}, \citenamefont {Ortega}, \citenamefont {Sanz}, \citenamefont
		{Berthelot}, \citenamefont {Garcia-Cordero}, \citenamefont {Renger},
		\citenamefont {Maerkl}, \citenamefont {Kreuzer},\ and\ \citenamefont
		{Quidant}}]{acimovic2014lspr}%
	\BibitemOpen
	\bibfield  {author} {\bibinfo {author} {\bibfnamefont {S.~S.}\ \bibnamefont
			{Acimovic}}, \bibinfo {author} {\bibfnamefont {M.~A.}\ \bibnamefont
			{Ortega}}, \bibinfo {author} {\bibfnamefont {V.}~\bibnamefont {Sanz}},
		\bibinfo {author} {\bibfnamefont {J.}~\bibnamefont {Berthelot}}, \bibinfo
		{author} {\bibfnamefont {J.~L.}\ \bibnamefont {Garcia-Cordero}}, \bibinfo
		{author} {\bibfnamefont {J.}~\bibnamefont {Renger}}, \bibinfo {author}
		{\bibfnamefont {S.~J.}\ \bibnamefont {Maerkl}}, \bibinfo {author}
		{\bibfnamefont {M.~P.}\ \bibnamefont {Kreuzer}}, \ and\ \bibinfo {author}
		{\bibfnamefont {R.}~\bibnamefont {Quidant}},\ }\href@noop {} {\bibfield
		{journal} {\bibinfo  {journal} {Nano letters}\ }\textbf {\bibinfo {volume}
			{14}},\ \bibinfo {pages} {2636} (\bibinfo {year} {2014})}\BibitemShut
	{NoStop}%
	\bibitem [{\citenamefont {Rodrigo}\ \emph {et~al.}(2015)\citenamefont
		{Rodrigo}, \citenamefont {Limaj}, \citenamefont {Janner}, \citenamefont
		{Etezadi}, \citenamefont {de~Abajo}, \citenamefont {Pruneri},\ and\
		\citenamefont {Altug}}]{rodrigo2015mid}%
	\BibitemOpen
	\bibfield  {author} {\bibinfo {author} {\bibfnamefont {D.}~\bibnamefont
			{Rodrigo}}, \bibinfo {author} {\bibfnamefont {O.}~\bibnamefont {Limaj}},
		\bibinfo {author} {\bibfnamefont {D.}~\bibnamefont {Janner}}, \bibinfo
		{author} {\bibfnamefont {D.}~\bibnamefont {Etezadi}}, \bibinfo {author}
		{\bibfnamefont {F.~J.~G.}\ \bibnamefont {de~Abajo}}, \bibinfo {author}
		{\bibfnamefont {V.}~\bibnamefont {Pruneri}}, \ and\ \bibinfo {author}
		{\bibfnamefont {H.}~\bibnamefont {Altug}},\ }\href@noop {} {\bibfield
		{journal} {\bibinfo  {journal} {Science}\ }\textbf {\bibinfo {volume}
			{349}},\ \bibinfo {pages} {165} (\bibinfo {year} {2015})}\BibitemShut
	{NoStop}%
	\bibitem [{\citenamefont {Green}\ and\ \citenamefont
		{Pillai}(2012)}]{green2012harnessing}%
	\BibitemOpen
	\bibfield  {author} {\bibinfo {author} {\bibfnamefont {M.~A.}\ \bibnamefont
			{Green}}\ and\ \bibinfo {author} {\bibfnamefont {S.}~\bibnamefont {Pillai}},\
	}\href@noop {} {\bibfield  {journal} {\bibinfo  {journal} {Nat. Photonics}\
		}\textbf {\bibinfo {volume} {6}},\ \bibinfo {pages} {130} (\bibinfo {year}
		{2012})}\BibitemShut {NoStop}%
	\bibitem [{\citenamefont {Reece}(2008)}]{reece2008plasmonics}%
	\BibitemOpen
	\bibfield  {author} {\bibinfo {author} {\bibfnamefont {P.~J.}\ \bibnamefont
			{Reece}},\ }\href@noop {} {\bibfield  {journal} {\bibinfo  {journal} {Nature
				Photonics}\ }\textbf {\bibinfo {volume} {2}},\ \bibinfo {pages} {333}
		(\bibinfo {year} {2008})}\BibitemShut {NoStop}%
	\bibitem [{\citenamefont {Juan}\ \emph {et~al.}(2011)\citenamefont {Juan},
		\citenamefont {Righini},\ and\ \citenamefont {Quidant}}]{juan2011plasmon}%
	\BibitemOpen
	\bibfield  {author} {\bibinfo {author} {\bibfnamefont {M.~L.}\ \bibnamefont
			{Juan}}, \bibinfo {author} {\bibfnamefont {M.}~\bibnamefont {Righini}}, \
		and\ \bibinfo {author} {\bibfnamefont {R.}~\bibnamefont {Quidant}},\
	}\href@noop {} {\bibfield  {journal} {\bibinfo  {journal} {Nat. Photonics}\
		}\textbf {\bibinfo {volume} {5}},\ \bibinfo {pages} {349} (\bibinfo {year}
		{2011})}\BibitemShut {NoStop}%
	\bibitem [{\citenamefont {Huidobro}\ \emph {et~al.}(2010)\citenamefont
		{Huidobro}, \citenamefont {Nesterov}, \citenamefont {Mart{\'\i}n-Moreno},\
		and\ \citenamefont {Garc{\'\i}a-Vidal}}]{huidobro2010transformation}%
	\BibitemOpen
	\bibfield  {author} {\bibinfo {author} {\bibfnamefont {P.~A.}\ \bibnamefont
			{Huidobro}}, \bibinfo {author} {\bibfnamefont {M.~L.}\ \bibnamefont
			{Nesterov}}, \bibinfo {author} {\bibfnamefont {L.}~\bibnamefont
			{Mart{\'\i}n-Moreno}}, \ and\ \bibinfo {author} {\bibfnamefont {F.~J.}\
			\bibnamefont {Garc{\'\i}a-Vidal}},\ }\href@noop {} {\bibfield  {journal}
		{\bibinfo  {journal} {NanoLetters}\ }\textbf {\bibinfo {volume} {10}},\
		\bibinfo {pages} {1985} (\bibinfo {year} {2010})}\BibitemShut {NoStop}%
	\bibitem [{\citenamefont {Pendry}\ \emph {et~al.}(2012)\citenamefont {Pendry},
		\citenamefont {Aubry}, \citenamefont {Smith},\ and\ \citenamefont
		{Maier}}]{pendry2012transformation}%
	\BibitemOpen
	\bibfield  {author} {\bibinfo {author} {\bibfnamefont {J.}~\bibnamefont
			{Pendry}}, \bibinfo {author} {\bibfnamefont {A.}~\bibnamefont {Aubry}},
		\bibinfo {author} {\bibfnamefont {D.}~\bibnamefont {Smith}}, \ and\ \bibinfo
		{author} {\bibfnamefont {S.}~\bibnamefont {Maier}},\ }\href@noop {}
	{\bibfield  {journal} {\bibinfo  {journal} {Science}\ }\textbf {\bibinfo
			{volume} {337}},\ \bibinfo {pages} {549} (\bibinfo {year}
		{2012})}\BibitemShut {NoStop}%
	\bibitem [{\citenamefont {Smith}\ \emph {et~al.}(2015)\citenamefont {Smith},
		\citenamefont {Stenger}, \citenamefont {Kristensen}, \citenamefont
		{Mortensen},\ and\ \citenamefont {Bozhevolnyi}}]{SmithRevCPP}%
	\BibitemOpen
	\bibfield  {author} {\bibinfo {author} {\bibfnamefont {C.~L.~C.}\
			\bibnamefont {Smith}}, \bibinfo {author} {\bibfnamefont {N.}~\bibnamefont
			{Stenger}}, \bibinfo {author} {\bibfnamefont {A.}~\bibnamefont {Kristensen}},
		\bibinfo {author} {\bibfnamefont {N.~A.}\ \bibnamefont {Mortensen}}, \ and\
		\bibinfo {author} {\bibfnamefont {S.~I.}\ \bibnamefont {Bozhevolnyi}},\
	}\href@noop {} {\bibfield  {journal} {\bibinfo  {journal} {Nanoscale}\
		}\textbf {\bibinfo {volume} {7}},\ \bibinfo {pages} {9355} (\bibinfo {year}
		{2015})}\BibitemShut {NoStop}%
	\bibitem [{\citenamefont {Nielsen}\ \emph {et~al.}(2008)\citenamefont
		{Nielsen}, \citenamefont {Fernandez-Cuesta}, \citenamefont {Boltasseva},
		\citenamefont {Volkov}, \citenamefont {Bozhevolnyi}, \citenamefont
		{Klukowska},\ and\ \citenamefont {Kristensen}}]{Nielsen:2008}%
	\BibitemOpen
	\bibfield  {author} {\bibinfo {author} {\bibfnamefont {R.~B.}\ \bibnamefont
			{Nielsen}}, \bibinfo {author} {\bibfnamefont {I.}~\bibnamefont
			{Fernandez-Cuesta}}, \bibinfo {author} {\bibfnamefont {A.}~\bibnamefont
			{Boltasseva}}, \bibinfo {author} {\bibfnamefont {V.~S.}\ \bibnamefont
			{Volkov}}, \bibinfo {author} {\bibfnamefont {S.~I.}\ \bibnamefont
			{Bozhevolnyi}}, \bibinfo {author} {\bibfnamefont {A.}~\bibnamefont
			{Klukowska}}, \ and\ \bibinfo {author} {\bibfnamefont {A.}~\bibnamefont
			{Kristensen}},\ }\href {\doibase 10.1364/OL.33.002800} {\bibfield  {journal}
		{\bibinfo  {journal} {Opt. Lett.}\ }\textbf {\bibinfo {volume} {33}},\
		\bibinfo {pages} {2800} (\bibinfo {year} {2008})}\BibitemShut {NoStop}%
	\bibitem [{\citenamefont {Gramotnev}\ and\ \citenamefont
		{Pile}(2004)}]{PileAPL}%
	\BibitemOpen
	\bibfield  {author} {\bibinfo {author} {\bibfnamefont {D.~K.}\ \bibnamefont
			{Gramotnev}}\ and\ \bibinfo {author} {\bibfnamefont {D.~F.~P.}\ \bibnamefont
			{Pile}},\ }\href {\doibase http://dx.doi.org/10.1063/1.1839283} {\bibfield
		{journal} {\bibinfo  {journal} {Appl. Phys. Lett.}\ }\textbf {\bibinfo
			{volume} {85}},\ \bibinfo {pages} {6323} (\bibinfo {year}
		{2004})}\BibitemShut {NoStop}%
	\bibitem [{\citenamefont {Jablan}\ \emph {et~al.}(2009)\citenamefont {Jablan},
		\citenamefont {Buljan},\ and\ \citenamefont
		{Solja{\v{c}}i{\'c}}}]{jablan2009plasmonics}%
	\BibitemOpen
	\bibfield  {author} {\bibinfo {author} {\bibfnamefont {M.}~\bibnamefont
			{Jablan}}, \bibinfo {author} {\bibfnamefont {H.}~\bibnamefont {Buljan}}, \
		and\ \bibinfo {author} {\bibfnamefont {M.}~\bibnamefont
			{Solja{\v{c}}i{\'c}}},\ }\href@noop {} {\bibfield  {journal} {\bibinfo
			{journal} {Physical review B}\ }\textbf {\bibinfo {volume} {80}},\ \bibinfo
		{pages} {245435} (\bibinfo {year} {2009})}\BibitemShut {NoStop}%
	\bibitem [{\citenamefont {Tassin}\ \emph {et~al.}(2012)\citenamefont {Tassin},
		\citenamefont {Koschny}, \citenamefont {Kafesaki},\ and\ \citenamefont
		{Soukoulis}}]{tassin2012comparison}%
	\BibitemOpen
	\bibfield  {author} {\bibinfo {author} {\bibfnamefont {P.}~\bibnamefont
			{Tassin}}, \bibinfo {author} {\bibfnamefont {T.}~\bibnamefont {Koschny}},
		\bibinfo {author} {\bibfnamefont {M.}~\bibnamefont {Kafesaki}}, \ and\
		\bibinfo {author} {\bibfnamefont {C.~M.}\ \bibnamefont {Soukoulis}},\
	}\href@noop {} {\bibfield  {journal} {\bibinfo  {journal} {Nature Photonics}\
		}\textbf {\bibinfo {volume} {6}},\ \bibinfo {pages} {259} (\bibinfo {year}
		{2012})}\BibitemShut {NoStop}%
	\bibitem [{\citenamefont {Wenger}\ \emph {et~al.}(2016)\citenamefont {Wenger},
		\citenamefont {Viola}, \citenamefont {Fogelstr{\"o}m}, \citenamefont
		{Tassin},\ and\ \citenamefont {Kinaret}}]{wenger2016optical}%
	\BibitemOpen
	\bibfield  {author} {\bibinfo {author} {\bibfnamefont {T.}~\bibnamefont
			{Wenger}}, \bibinfo {author} {\bibfnamefont {G.}~\bibnamefont {Viola}},
		\bibinfo {author} {\bibfnamefont {M.}~\bibnamefont {Fogelstr{\"o}m}},
		\bibinfo {author} {\bibfnamefont {P.}~\bibnamefont {Tassin}}, \ and\ \bibinfo
		{author} {\bibfnamefont {J.}~\bibnamefont {Kinaret}},\ }\href@noop {}
	{\bibfield  {journal} {\bibinfo  {journal} {Physical Review B}\ }\textbf
		{\bibinfo {volume} {94}},\ \bibinfo {pages} {205419} (\bibinfo {year}
		{2016})}\BibitemShut {NoStop}%
	\bibitem [{\citenamefont {Xiao}\ \emph {et~al.}(2016)\citenamefont {Xiao},
		\citenamefont {Zhu}, \citenamefont {Li},\ and\ \citenamefont
		{Mortensen}}]{Xiao2016}%
	\BibitemOpen
	\bibfield  {author} {\bibinfo {author} {\bibfnamefont {S.}~\bibnamefont
			{Xiao}}, \bibinfo {author} {\bibfnamefont {X.}~\bibnamefont {Zhu}}, \bibinfo
		{author} {\bibfnamefont {B.-H.}\ \bibnamefont {Li}}, \ and\ \bibinfo {author}
		{\bibfnamefont {N.~A.}\ \bibnamefont {Mortensen}},\ }\href {\doibase
		10.1007/s11467-016-0551-z} {\bibfield  {journal} {\bibinfo  {journal} {Front.
				Phys.}\ }\textbf {\bibinfo {volume} {11}},\ \bibinfo {pages} {117801}
		(\bibinfo {year} {2016})}\BibitemShut {NoStop}%
	\bibitem [{\citenamefont {Garc\'ia~de Abajo}(2014)}]{AbajoACSP}%
	\BibitemOpen
	\bibfield  {author} {\bibinfo {author} {\bibfnamefont {F.~J.}\ \bibnamefont
			{Garc\'ia~de Abajo}},\ }\href@noop {} {\bibfield  {journal} {\bibinfo
			{journal} {ACS Photonics}\ }\textbf {\bibinfo {volume} {1}},\ \bibinfo
		{pages} {135} (\bibinfo {year} {2014})}\BibitemShut {NoStop}%
	\bibitem [{\citenamefont {Gon{\c{c}}alves}\ \emph
		{et~al.}(2017{\natexlab{a}})\citenamefont {Gon{\c{c}}alves}, \citenamefont
		{Xiao}, \citenamefont {Peres},\ and\ \citenamefont
		{Mortensen}}]{gonccalves2017hybridized}%
	\BibitemOpen
	\bibfield  {author} {\bibinfo {author} {\bibfnamefont {P.~A.~D.}\
			\bibnamefont {Gon{\c{c}}alves}}, \bibinfo {author} {\bibfnamefont
			{S.}~\bibnamefont {Xiao}}, \bibinfo {author} {\bibfnamefont {N.}~\bibnamefont
			{Peres}}, \ and\ \bibinfo {author} {\bibfnamefont {N.~A.}\ \bibnamefont
			{Mortensen}},\ }\href@noop {} {\bibfield  {journal} {\bibinfo  {journal} {ACS
				Photonics}\ } (\bibinfo {year} {2017}{\natexlab{a}})}\BibitemShut {NoStop}%
	\bibitem [{\citenamefont {Liu}\ \emph {et~al.}(2013)\citenamefont {Liu},
		\citenamefont {Zhang}, \citenamefont {Ma}, \citenamefont {Cai}, \citenamefont
		{Wang},\ and\ \citenamefont {Xu}}]{Liu:13}%
	\BibitemOpen
	\bibfield  {author} {\bibinfo {author} {\bibfnamefont {P.}~\bibnamefont
			{Liu}}, \bibinfo {author} {\bibfnamefont {X.}~\bibnamefont {Zhang}}, \bibinfo
		{author} {\bibfnamefont {Z.}~\bibnamefont {Ma}}, \bibinfo {author}
		{\bibfnamefont {W.}~\bibnamefont {Cai}}, \bibinfo {author} {\bibfnamefont
			{L.}~\bibnamefont {Wang}}, \ and\ \bibinfo {author} {\bibfnamefont
			{J.}~\bibnamefont {Xu}},\ }\href@noop {} {\bibfield  {journal} {\bibinfo
			{journal} {Opt. Express}\ }\textbf {\bibinfo {volume} {21}},\ \bibinfo
		{pages} {32432} (\bibinfo {year} {2013})}\BibitemShut {NoStop}%
	\bibitem [{\citenamefont {Smirnova}\ \emph {et~al.}(2016)\citenamefont
		{Smirnova}, \citenamefont {Mousavi}, \citenamefont {Wang}, \citenamefont
		{Kivshar},\ and\ \citenamefont {Khanikaev}}]{Smirnova:2016}%
	\BibitemOpen
	\bibfield  {author} {\bibinfo {author} {\bibfnamefont {D.}~\bibnamefont
			{Smirnova}}, \bibinfo {author} {\bibfnamefont {S.~H.}\ \bibnamefont
			{Mousavi}}, \bibinfo {author} {\bibfnamefont {Z.}~\bibnamefont {Wang}},
		\bibinfo {author} {\bibfnamefont {Y.~S.}\ \bibnamefont {Kivshar}}, \ and\
		\bibinfo {author} {\bibfnamefont {A.~B.}\ \bibnamefont {Khanikaev}},\ }\href
	{\doibase 10.1021/acsphotonics.6b00116} {\bibfield  {journal} {\bibinfo
			{journal} {ACS Photonics}\ }\textbf {\bibinfo {volume} {3}},\ \bibinfo
		{pages} {875} (\bibinfo {year} {2016})}\BibitemShut {NoStop}%
	\bibitem [{\citenamefont {Gon{\c{c}}alves}\ \emph
		{et~al.}(2017{\natexlab{b}})\citenamefont {Gon{\c{c}}alves}, \citenamefont
		{Bozhevolnyi}, \citenamefont {Mortensen},\ and\ \citenamefont
		{Peres}}]{gonccalves2017universal}%
	\BibitemOpen
	\bibfield  {author} {\bibinfo {author} {\bibfnamefont {P.~A.~D.}\
			\bibnamefont {Gon{\c{c}}alves}}, \bibinfo {author} {\bibfnamefont {S.~I.}\
			\bibnamefont {Bozhevolnyi}}, \bibinfo {author} {\bibfnamefont {N.~A.}\
			\bibnamefont {Mortensen}}, \ and\ \bibinfo {author} {\bibfnamefont
			{N.}~\bibnamefont {Peres}},\ }\href@noop {} {\bibfield  {journal} {\bibinfo
			{journal} {Optica}\ }\textbf {\bibinfo {volume} {4}},\ \bibinfo {pages} {595}
		(\bibinfo {year} {2017}{\natexlab{b}})}\BibitemShut {NoStop}%
	\bibitem [{\citenamefont {Maradudin}\ and\ \citenamefont
		{Visscher}(1985)}]{maradudin1985electrostatic}%
	\BibitemOpen
	\bibfield  {author} {\bibinfo {author} {\bibfnamefont {A.}~\bibnamefont
			{Maradudin}}\ and\ \bibinfo {author} {\bibfnamefont {W.}~\bibnamefont
			{Visscher}},\ }\href@noop {} {\bibfield  {journal} {\bibinfo  {journal} {Z.
				Phys. B Cond. Matter}\ }\textbf {\bibinfo {volume} {60}},\ \bibinfo {pages}
		{215} (\bibinfo {year} {1985})}\BibitemShut {NoStop}%
	\bibitem [{\citenamefont {Sturman}\ \emph {et~al.}(2014)\citenamefont
		{Sturman}, \citenamefont {Podivilov},\ and\ \citenamefont
		{Gorkunov}}]{sturman2014plasmons}%
	\BibitemOpen
	\bibfield  {author} {\bibinfo {author} {\bibfnamefont {B.}~\bibnamefont
			{Sturman}}, \bibinfo {author} {\bibfnamefont {E.}~\bibnamefont {Podivilov}},
		\ and\ \bibinfo {author} {\bibfnamefont {M.}~\bibnamefont {Gorkunov}},\
	}\href@noop {} {\bibfield  {journal} {\bibinfo  {journal} {JOSA B}\ }\textbf
		{\bibinfo {volume} {31}},\ \bibinfo {pages} {1607} (\bibinfo {year}
		{2014})}\BibitemShut {NoStop}%
	\bibitem [{\citenamefont {Mayergoyz}\ \emph {et~al.}(2005)\citenamefont
		{Mayergoyz}, \citenamefont {Fredkin},\ and\ \citenamefont
		{Zhang}}]{mayergoyz2005electrostatic}%
	\BibitemOpen
	\bibfield  {author} {\bibinfo {author} {\bibfnamefont {I.~D.}\ \bibnamefont
			{Mayergoyz}}, \bibinfo {author} {\bibfnamefont {D.~R.}\ \bibnamefont
			{Fredkin}}, \ and\ \bibinfo {author} {\bibfnamefont {Z.}~\bibnamefont
			{Zhang}},\ }\href@noop {} {\bibfield  {journal} {\bibinfo  {journal} {Phys.
				Rev. B}\ }\textbf {\bibinfo {volume} {72}},\ \bibinfo {pages} {155412}
		(\bibinfo {year} {2005})}\BibitemShut {NoStop}%
	\bibitem [{\citenamefont {Arias}\ and\ \citenamefont
		{Maradudin}(2013)}]{arias2013scattering}%
	\BibitemOpen
	\bibfield  {author} {\bibinfo {author} {\bibfnamefont {R.~E.}\ \bibnamefont
			{Arias}}\ and\ \bibinfo {author} {\bibfnamefont {A.~A.}\ \bibnamefont
			{Maradudin}},\ }\href@noop {} {\bibfield  {journal} {\bibinfo  {journal}
			{Optics express}\ }\textbf {\bibinfo {volume} {21}},\ \bibinfo {pages} {9734}
		(\bibinfo {year} {2013})}\BibitemShut {NoStop}%
	\bibitem [{\citenamefont {Zayats}\ \emph {et~al.}(2005)\citenamefont {Zayats},
		\citenamefont {Smolyaninov},\ and\ \citenamefont
		{Maradudin}}]{zayats2005nano}%
	\BibitemOpen
	\bibfield  {author} {\bibinfo {author} {\bibfnamefont {A.~V.}\ \bibnamefont
			{Zayats}}, \bibinfo {author} {\bibfnamefont {I.~I.}\ \bibnamefont
			{Smolyaninov}}, \ and\ \bibinfo {author} {\bibfnamefont {A.~A.}\ \bibnamefont
			{Maradudin}},\ }\href@noop {} {\bibfield  {journal} {\bibinfo  {journal}
			{Phys. Rep.}\ }\textbf {\bibinfo {volume} {408}},\ \bibinfo {pages} {131}
		(\bibinfo {year} {2005})}\BibitemShut {NoStop}%
	\bibitem [{\citenamefont {Raether}(1988)}]{raether1988surface}%
	\BibitemOpen
	\bibfield  {author} {\bibinfo {author} {\bibfnamefont {H.}~\bibnamefont
			{Raether}},\ }in\ \href@noop {} {\emph {\bibinfo {booktitle} {Surface
				plasmons on smooth and rough surfaces and on gratings}}}\ (\bibinfo
	{publisher} {Springer},\ \bibinfo {year} {1988})\ pp.\ \bibinfo {pages}
	{4--39}\BibitemShut {NoStop}%
	\bibitem [{\citenamefont {Pereira}\ \emph {et~al.}(2003)\citenamefont
		{Pereira}, \citenamefont {Farias},\ and\ \citenamefont
		{Costa~Filho}}]{pereira2003surface}%
	\BibitemOpen
	\bibfield  {author} {\bibinfo {author} {\bibfnamefont {J.~M.}\ \bibnamefont
			{Pereira}}, \bibinfo {author} {\bibfnamefont {G.}~\bibnamefont {Farias}}, \
		and\ \bibinfo {author} {\bibfnamefont {R.}~\bibnamefont {Costa~Filho}},\
	}\href@noop {} {\bibfield  {journal} {\bibinfo  {journal} {Eur. Phys. J. B.}\
		}\textbf {\bibinfo {volume} {36}},\ \bibinfo {pages} {137} (\bibinfo {year}
		{2003})}\BibitemShut {NoStop}%
	\bibitem [{\citenamefont {Chubchev}\ \emph {et~al.}(2017)\citenamefont
		{Chubchev}, \citenamefont {Nechepurenko}, \citenamefont {Dorofeenko},
		\citenamefont {Vinogradov},\ and\ \citenamefont
		{Lisyansky}}]{chubchev2017surface}%
	\BibitemOpen
	\bibfield  {author} {\bibinfo {author} {\bibfnamefont {E.}~\bibnamefont
			{Chubchev}}, \bibinfo {author} {\bibfnamefont {I.}~\bibnamefont
			{Nechepurenko}}, \bibinfo {author} {\bibfnamefont {A.}~\bibnamefont
			{Dorofeenko}}, \bibinfo {author} {\bibfnamefont {A.}~\bibnamefont
			{Vinogradov}}, \ and\ \bibinfo {author} {\bibfnamefont {A.}~\bibnamefont
			{Lisyansky}},\ }\href@noop {} {\bibfield  {journal} {\bibinfo  {journal}
			{arXiv preprint arXiv:1711.08115}\ } (\bibinfo {year} {2017})}\BibitemShut
	{NoStop}%
	\bibitem [{\citenamefont {Garc\'ia~de Abajo}(2013)}]{garcia2013multiple}%
	\BibitemOpen
	\bibfield  {author} {\bibinfo {author} {\bibfnamefont {F.~J.}\ \bibnamefont
			{Garc\'ia~de Abajo}},\ }\href@noop {} {\bibfield  {journal} {\bibinfo
			{journal} {ACS nano}\ }\textbf {\bibinfo {volume} {7}},\ \bibinfo {pages}
		{11409} (\bibinfo {year} {2013})}\BibitemShut {NoStop}%
	\bibitem [{\citenamefont {Kumar}\ \emph {et~al.}(2015)\citenamefont {Kumar},
		\citenamefont {Low}, \citenamefont {Fung}, \citenamefont {Avouris},\ and\
		\citenamefont {Fang}}]{kumar2015tunable}%
	\BibitemOpen
	\bibfield  {author} {\bibinfo {author} {\bibfnamefont {A.}~\bibnamefont
			{Kumar}}, \bibinfo {author} {\bibfnamefont {T.}~\bibnamefont {Low}}, \bibinfo
		{author} {\bibfnamefont {K.~H.}\ \bibnamefont {Fung}}, \bibinfo {author}
		{\bibfnamefont {P.}~\bibnamefont {Avouris}}, \ and\ \bibinfo {author}
		{\bibfnamefont {N.~X.}\ \bibnamefont {Fang}},\ }\href@noop {} {\bibfield
		{journal} {\bibinfo  {journal} {Nano letters}\ }\textbf {\bibinfo {volume}
			{15}},\ \bibinfo {pages} {3172} (\bibinfo {year} {2015})}\BibitemShut
	{NoStop}%
	\bibitem [{\citenamefont {Bart}\ and\ \citenamefont
		{Warnock}(1973)}]{Bart1973}%
	\BibitemOpen
	\bibfield  {author} {\bibinfo {author} {\bibfnamefont {G.}~\bibnamefont
			{Bart}}\ and\ \bibinfo {author} {\bibfnamefont {R.}~\bibnamefont {Warnock}},\
	}\href@noop {} {\bibfield  {journal} {\bibinfo  {journal} {SIAM J. on Math.
				Anal.}\ }\textbf {\bibinfo {volume} {4}},\ \bibinfo {pages} {609} (\bibinfo
		{year} {1973})}\BibitemShut {NoStop}%
\end{thebibliography}

%

\end{document}